\documentclass[12pt,preprint]{aastex}

\slugcomment{To appear in Astrophysical Journal}

\shorttitle{Deep \textit{Chandra} observation of XDCP J0044.0-2033}
\shortauthors{Tozzi P.  et al.}

\usepackage[caption=false]{subfig}
\usepackage{amsmath}
\usepackage{times}
\usepackage{natbib}
\usepackage[colorlinks,linkcolor=blue,anchorcolor=blue,citecolor=blue]{hyperref}
\usepackage[hyperref]{backref}
\usepackage{txfonts}
\usepackage{graphicx}
\usepackage{color}
\usepackage{longtable,lscape}

\newcommand{\mytilde}{\raise.19ex\hbox{$\scriptstyle\sim$}}

\begin{document}

\title{\textit{Chandra} deep observation of XDCP J0044.0-2033, a massive
  galaxy cluster at $z > 1.5$}


\author{P. Tozzi\altaffilmark{1}, J. S. Santos\altaffilmark{1}, M.  J.  Jee\altaffilmark{2}, R. Fassbender\altaffilmark{3}, P. Rosati\altaffilmark{4,1}, A. Nastasi\altaffilmark{5}, W. Forman\altaffilmark{6},   B. Sartoris\altaffilmark{7}, S. Borgani\altaffilmark{7,8}, H. Boehringer\altaffilmark{9}, B. Altieri\altaffilmark{10}, G. W. Pratt\altaffilmark{11},  M. Nonino\altaffilmark{8}, C. Jones\altaffilmark{6}}

\altaffiltext{1}{INAF - Osservatorio Astrofisico di Arcetri, Largo E. Fermi 5, 
50125, Firenze, Italy}
\altaffiltext{2}{Department of Physics , University of California, Davis
One Shields Avenue, Davis, CA 95616-8677}
\altaffiltext{3}{INAF - Osservatorio Astronomico di Roma (OAR), via Frascati 33, Monte Porzio Catone, Italy}  
\altaffiltext{4}{ Universit\`a degli Studi di Ferrara, Via Savonarola, 9 - 44121 Ferrara , Italy}
\altaffiltext{5}{Istitut d'Astrophysique Spatiale, CNRS, Bat. 121, Universit\'e Paris-Sud, Orsay, France}
\altaffiltext{6}{Harvard-Smithsonian Center for Astrophysics 60 Garden Street, Cambridge, MA 02138, USA}
\altaffiltext{7}{Universit\`a degli Studi di Trieste, Dipartimento di Fisica, Via A.Valerio, 2 - 34127 Trieste, Italy}  
\altaffiltext{8}{INAF - Osservatorio Astronomico di Trieste, via G. B. Tiepolo 11, 34143 Trieste, Italy}  
\altaffiltext{9}{Max-Planck-Institut für extraterrestrische Physik Giessenbachstr.1 , D-85748 Garching , Germany }
\altaffiltext{10}{European Space Astronomy Centre (ESAC), European Space Agency, Apartado de Correos
78, 28691 Villanueva de la Canada, Madrid, Spain} 
\altaffiltext{11}{CEA Saclay, Service d’Astrophysique, L’Orme des Merisiers, Bat. 709, 91191 Gif-sur-Yvette Cedex, France}




\begin{abstract}
We report the analysis of the Chandra observation of XDCP J0044.0-2033, a massive,
distant ($z = 1.579$) galaxy cluster discovered in the XDCP survey.
The total exposure time of 380 ks with {\it Chandra} ACIS-S provides the deepest
X-ray observation currently achieved on a massive, high redshift cluster.  
Extended emission from the Intra Cluster Medium (ICM)
is detected at a very high significance level ($S/N \sim 20$) on a circular 
region with a $44"$ radius, corresponding to $R_{ext}=375$ kpc at the cluster redshift.  We
perform an X-ray spectral fit of the ICM emission modeling the
spectrum with a single-temperature thermal \textit{mekal} model.  Our analysis provides a global
temperature $kT=6.7^{+1.3}_{-0.9}$ keV, and a iron abundance $Z_{Fe} =
0.41_{-0.26}^{+0.29}Z_{Fe_\odot}$ (error bars correspond to 1 $\sigma$).  We fit the background-subtracted
surface brightness profile with a single $beta$-model out to $44"$, finding a rather flat profile with no hints of
a cool core.   We derive the deprojected electron density profile and compute the  ICM 
mass within the extraction radius $R_{ext}=375$ kpc to be $M_{ICM}(r<R_{ext}) = (1.48 \pm 0.20) \times 10^{13} M_\odot$.  Under the
assumption of hydrostatic equilibrium and assuming isothermality within $R_{ext}$, the total mass is $M_{2500}=
1.23_{-0.27}^{+0.46} \times 10 ^{14} M_\odot$ for $R_{2500} = 240_{-20}^{+30}$
kpc.   Extrapolating the profile at radii larger than the extraction radius $R_{ext}$ we find $M_{500} =
3.2_{-0.6}^{+0.9} \times 10 ^{14}M_\odot$ for $R_{500} = 562_{-37}^{+50}$ kpc.  This analysis establishes the existence of virialized, massive galaxy clusters
at redshift $z\sim 1.6$, paving the way to the investigation of the progenitors of the most massive clusters today. 
Given its mass and the XDCP survey volume, XDCP J0044.0-2033 does not create significant tension with the 
WMAP-7 $\Lambda$CDM cosmology.
\end{abstract}





  


\section{Introduction}

The search and characterization of distant ($z>1$) clusters of galaxies has been a major field of research in extragalactic astronomy
in the last ten years.  A systematic investigation of galaxy clusters at high redshift can provide at the same time strong constraints on
cosmological parameters, on the physics of the Intra Cluster Medium (ICM) and of the interaction of the ICM with cluster galaxies, and on
the evolution of the cluster galaxy population.  Over the last five years, the number of known clusters at redshift $z>1$ has dramatically
increased from a few to several tens, while their characterization still remains very challenging.  In addition, it is difficult to
assemble a sample of high-z clusters with a well-defined selection function, given the wide range of detection techniques used to find the clusters.  

At present, there is an increasing effort to build  homogeneous and well 
characterized samples of high-z clusters based on several complementary approaches.
X-ray, infrared (IR) and Sunyaev-Zeldovich (SZ) surveys are the main tools used to build 
cluster samples.  X-ray surveys play a key role in this context.   Thanks
to the X-ray thermal emission from the ICM, which is the largest
baryonic component of galaxy clusters, it is possible to identify
galaxy clusters as X-ray extended sources up to $z>1$.  This was shown
already with the ROSAT satellite in the ROSAT Deep Cluster Survey
\citet{2002Rosati}, where several massive clusters were found up to $z
\sim 1.3$ \citep{2001Stanford,2002Stanford,2004Rosati}.  More
recently, the XMM-Newton Distant Cluster Project \citep[XDCP,][]{2011bFassbender} exploited the capability of XMM in identifying high-z cluster candidates, 
thanks to its large throughput and field of view.  The XDCP has proven to be particularly efficient with
$47$\footnote{Among the 47 XDCP clusters at $z>0.8$, 26 are published in the quoted papers, 
while the remaining 21 spectroscopically confirmed
clusters are still unpublished.} clusters {\sl spectroscopically} confirmed to be at $0.8<z<1.6$
\citep{2005Mullis,2009Santos,2011Nastasi,2011bFassbender,2012Pierini,2014Nastasi}, nearly five times the RDCS sample at $z>0.8$.  Other
serendipitous surveys based on the archival data of {\sl Chandra},
XMM-Newton and Swift-XRT archives are ongoing \citep[see][]{2006champs,2011bFassbender,clerc12,2014Tozzi}, together with fewer
dedicated, contiguous survey \citep{2007Finoguenov,2007XMMLSS}.  For a recent (upated to 2012) summary of the current X-ray cluster surveys see \citet{2012Tundo}.

IR and SZ surveys are contributing significantly to high redshift cluster detections 
\citep{2010Demarco,2011Foley,2011Brodwin,2012Brodwin,2012Stanford,2014Rettura} and will contribute important samples in the
future, particularly due to the lack of a timely, wide-angle and sensitive X-ray survey.  Nevertheless, the X-ray band presently provides the best diagnostics of 
the dynamical state of the clusters, of the thermo- and chemo-dynamical properties of the ICM and a robust measurement of cluster masses.
This can be achieved through the spectral analysis of the ICM emission which provides the mass of
the gas and its temperature.  Both quantities can be combined to obtain the integrated pseudo-pressure
parameter $Y_X \equiv T_X \times M_{ICM}$, considered a robust mass proxy within $R_{500}$, as shown by numerical simulations
\citep{kravtsov06}.  Moreover, if the data are sufficiently deep for a spatially resolved
analysis, the hydrostatic equilibrium equation can be directly applied to derive a full mass profile.
SZ observation are also a direct probe of the ICM, however the capability of SZ data to characterize the thermodynamics of the ICM is still below that of X-ray data.  
As of today, the calibration of the SZ mass proxy still relies on a cross-calibration with X-ray
data and weak lensing measurements \citep{2011PlanckXI,andersson2012,2013PlanckIII,2014vonderLinden,2014Donahue} or dynamical mass estimates \citep{2013Sifon}.  
Finally, the information that can be derived from optical and IR observations of distant clusters depends critically on how many galaxies are sampled. 
They give sparse information on the dynamical status and virial mass estimates are strongly dependent on the number of galaxies that are observed. 

Up to now, the community has been more interested in pushing the limit towards 
the most distant clusters, rather than in assembling a well characterized 
high-z sample.  If we focus on the most distant cluster candidates, we find that only
nine clusters have been spectroscopically confirmed at $z\ge 1.5$ to
date and only some of them have estimated masses in excess of $10^{14}$ $M_\odot$
\citep{2011Brodwin,2012Brodwin,2011Fassbender,2011Nastasi,2011Santos,2012Zeimann,2012Stanford}.
We note that most of these high redshift clusters have been identified in IR
surveys or by combining near infrared (NIR) and X-ray data\footnote{The most
  distant confirmed SZ-selected clusters to date are reported at $z\sim 1.32$ \citep{2012Stalder} and $z=1.478$   \citep{2014Bayliss}.}.  In the last three years a moderate investment
of \textit{Chandra} time has been dedicated to the X-ray follow-up observation of some of these clusters at $z>1.5$.  However, no
prominent, X-ray diffuse emission has been detected, so that no
constraints on the cluster dynamical state could be derived \citep{2010Papovich,2011Gobat,2012Pierre}.  The lack of extended
X-ray emission in these systems suggests low X-ray luminosities and points towards a re-classification of these objects as protoclusters.
Instead, X--ray emission has been detected in IR-selected clusters at $z> 1$ but, again, their low X--ray flux did not allow for a
detailed analysis \citep{2006Stanford,2009Andreon,2011Brodwin}.  Only recently, a serendipitously discovered cluster in a
deep, ACIS-S {\it Chandra} observation, allowed us to measure the X-ray redshift and the temperature of a $z\sim 1.5$ cluster \citep{2013Tozzi}.

In this scenario, deep \textit{Chandra} observations of {\sl bona fide} X-ray selected clusters are the only available means to investigate
cluster physics at very high-z \citep[see][]{2002Rosati,2013bTozzi}.  These observations are time-expensive but also extremely valuable, 
since they allow us to peer into an epoch, at a lookback time larger than $9$ Gyr ($z>1.5$), which recent observations indicate is key
for the assembly of cluster mass and the inversion of the star-formation--density relation in cluster galaxies.  Given the large amount of time
needed to observe clusters at such a high redshift, a crucial
requirement is to have an X-ray selected target, whose diffuse X-ray 
emission has been unambiguously assessed.  As matter of fact, any deep {\sl Chandra} observation 
of a distant, {\sl bona fide} cluster is providing a secure scientific
return which will be unrivaled until the next generation of
high-angular resolution X-ray satellites\footnote{At present, the only future mission which foresees an improvement 
with respect to the {\it Chandra} optics is SMART-X, see http://smart-x.cfa.harvard.edu/doc/2011-10-smartx-rfi-response.pdf.  
The planned mission Athena \citep{2012Barcons,2013Nandra} will have a much larger effective area on a significantly larger field of view, however, the current 
requirements on the angular resolution correspond at best to $5"$ half energy width (HEW, see http://www.the-athena-x-ray-observatory.eu/).}. 
This strategy has been shown to be highly successful for XMMU J2235.3-2557 observed for 200 ks with \textit{Chandra} ACIS-S \citep{2009Rosati}, and whose accurate mass estimate has
triggered a large number of theoretical speculations on its cosmological implications \citep[e.g.,][]{2010Sartoris,2011Mortonson,2012Harrison}.  

In this paper we present the deep, 380 ks {\it Chandra} observation of XDCP J0044.0-2033 (hereafter XDCP0044) at
$z=1.58$ \citep{2011Santos}, the X-ray selected, distant cluster with the largest estimated mass at $z>1.5$.  XDCP0044 was discovered
serendipitously in the XMM-Newton data in the XDCP.  The XMM discovery data,
however, allows only a robust measure of the X--ray luminosity of the most distant clusters \citep[]{2011bFassbender} and only
fairly uncertain masses, simply based on scaling relations, can be derived.  The {\sl Chandra} data presented in this work
constitutes the deepest X-ray observation to date on a cluster at  $z>1.0$.

The paper is organized as follows.  In \S 2 we describe the discovery data and give a brief summary of the available multi-wavelength data
set.  Then we present our X-ray data reduction procedure and spectral analysis, which provides us with global temperature and
iron abundance.  We also investigate the surface brightness properties of the
ICM.  In \S 3 we derive the ICM mass and the total cluster mass for a given density contrast with respect to the critical value.  In \S 4 we
discuss the systematics on the mass measurements, the cosmological implications, the possible presence of substructures in the ICM, 
and briefly discuss the future of high-z cluster studies in the light of planned X-ray missions.  Finally, our conclusions are summarized in \S 5.
Throughout the paper, we adopt the seven-year WMAP cosmology
($\Omega_{\Lambda} =0.73 $, $\Omega_m =0.27$, and $H_0 = 70.4 $ km
s$^{-1}$ Mpc$^{-1}$ \citep{2011Komatsu}.  In this cosmology, $1
\arcsec$ on the sky corresponds to $8.554$ kpc at $z = 1.58$.  Quoted
errors always correspond to a 1 $\sigma$ confidence level.

\section{Cluster identification and X-ray data analysis}

\subsection{Discovery data and multi-wavelength data set}

XDCP0044 was serendipitously detected in the archive of the 
X-ray observatory XMM-Newton by the XDCP, as an extended source at 5$\sigma$ c.l. at an 
off--axis angle of $10.8'$ and J2000 coordinates RA$=00$h$44$m$05.2$s, DEC$=-20^\circ 33^\prime 59.7^{\prime\prime}$, with an effective exposure time of $8.5$ ksec.  The unabsorbed flux is
measured from XMM-Newton data to be $S_{0.5-2.0} =(1.6 \pm 0.3) \times
10^{-14}$ erg s$^{-1}$ cm$^{-2}$ based on the growth curve analysis, within a radius of $35''$, 
corresponding to $296$ kpc at $z=1.58$.  

\begin{figure}
\begin{center}
\includegraphics[width=14cm,angle=0]{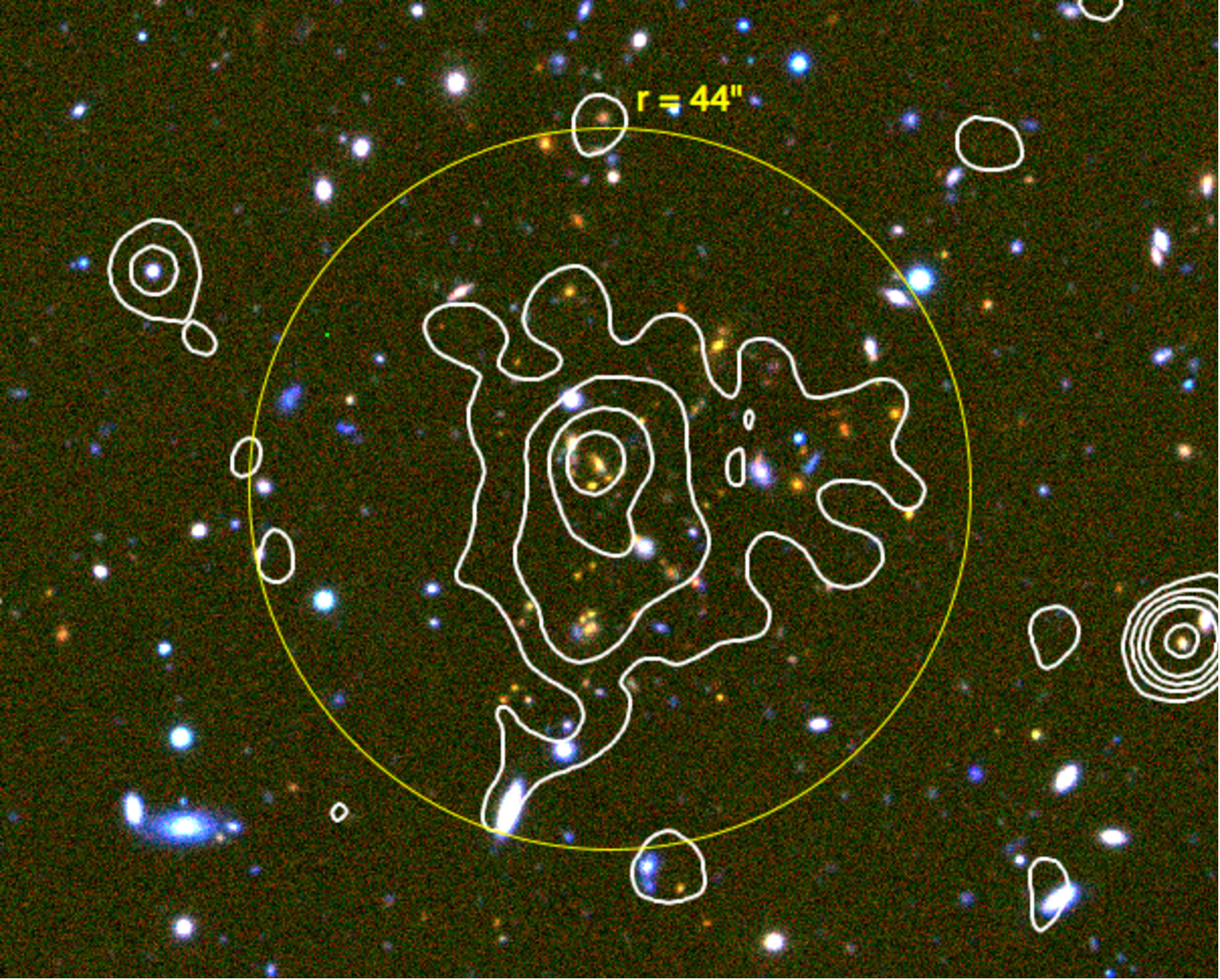} 
\end{center}
\caption{\label{color_image}Optical IJKs color image of XDCP0044 with {\it Chandra} smoothed soft-band contours overlaid.
Contours correspond to levels of 0.11, 0.3, 0.6 and 1.0 counts per pixel, to be compared with a background
level of $3.5 \times 10^{-2}$ cts per pixel in the original image (1 pixel = $0.492"$).
The image is obtained from Subaru/Suprime-Cam (V and i bands) and Hawk-I at VLT (J and Ks band) and has a size of $ 2.5 \arcmin \times 2 \arcmin$.   The solid circle has a radius of
 $44 \arcsec$ (corresponding to $375$ kpc at $z=1.58$), and shows the region used for the X-ray spectral analysis.  }
\end{figure}

We detected a galaxy overdensity associated with the diffuse X-ray emission using medium-deep imaging in I- and H-bands 
from EFOSC and SOFI respectively, at ESO/NTT.  The color image with X-ray contours overlaid is shown in Figure \ref{color_image}.  Subsequent optical 
spectroscopy from VLT/FORS2 later confirmed the cluster redshift by identifying three cluster members \citep{2011Santos}.  
Recently, new VLT/FORS2 spectroscopy secured other four cluster members, bringing the total to seven already published in the literature
\citep{2014Fassbender}.  Most recent progress in optical spectroscopy added at least five new secure cluster
members, bringing the total number of confirmed members to twelve (Nastasi et al. in preparation).  Among the
twelve confirmed members, six are within $20\arcsec$ from the X-ray center, and therefore are embedded in the ICM extended emission.

Since the discovery of XDCP0044, we have embarked on a demanding observational campaign, given the uniqueness of this system. Currently, our multi-$\lambda$ dataset 
includes deep J/Ks imaging from VLT/Hawk-I presented in \citet{2014Fassbender}, additional VLT/FORS2 data providing the twelve confirmed clusters members 
previously mentioned, recent VLT/KMOS infrared IFU spectroscopy which will augment our sample of cluster 
members, as well as mid-to-far infrared imaging from \textit{Spitzer} (Verdugo et al. in preparation) and \textit{Herschel} 
(Santos et al., in preparation). In addition to our dedicated programs, we also have access to deep, wide-field imaging in I and V bands 
from Subaru/\textit{Suprime-Cam}.  The analysis of this data, to be published in forthcoming papers, 
will provide a detailed characterization of the galaxy population of this massive, high-$z$ cluster, giving important 
clues on galaxy evolutionary processes, such as the amount and distribution of star formation in the cluster members.

\subsection{X-ray data reduction}

XDCP0044 was observed with a \textit{Chandra} Large Program observation of 380 ks with ACIS-S
granted in Cycle 14 (PI P. Tozzi).  The observations were completed in
the period October-December 2013.  The nominal total exposure time excluding the dead-time correction
(corresponding to the {\tt LIVETIME} keyword in the header of {\sl{Chandra}} fits files) amounts to 371.6 ks.

We performed a standard data reduction starting from the level=1 event
files, using the {\tt CIAO 4.6} software package, with the most recent
version of the Calibration Database ({\tt CALDB 4.6.3}).  Since our
observation is taken in the VFAINT mode we ran the task {\tt
  acis$\_$process$\_$events} to flag background events that are most
likely associated with cosmic rays and distinguish them from real
X-ray events. With this procedure, the ACIS particle background can be
significantly reduced compared to the standard grade selection.  The
data is filtered to include only the standard event grades 0, 2, 3, 4
and 6.  We checked visually for hot columns left after the standard
reduction.  For exposures taken in VFAINT mode, there are practically
no hot columns or flickering pixels left after filtering out bad events. We finally
filter time intervals with high background by performing a $3\sigma$
clipping of the background level using the script {\tt
  analyze\_ltcrv}.  The final useful exposure time amounts to 366.8 ks.  The
removed time intervals therefore amount to 4.8 ks, about 1.3\% of the
nominal exposure time.  We remark that our spectral analysis is not
affected by any possible undetected flare, since we are able to compute
the background in the same observation from a large, source-free region around the cluster, thus
taking into account any possible spectral distortion of the background
itself induced by unnoticed flares.  Eventually, we will explore the dependence of our results
on the background choice, including also a synthetic background.


\subsection{Point source removal and aperture photometry}

\begin{figure}
\includegraphics[width=10cm,angle=0]{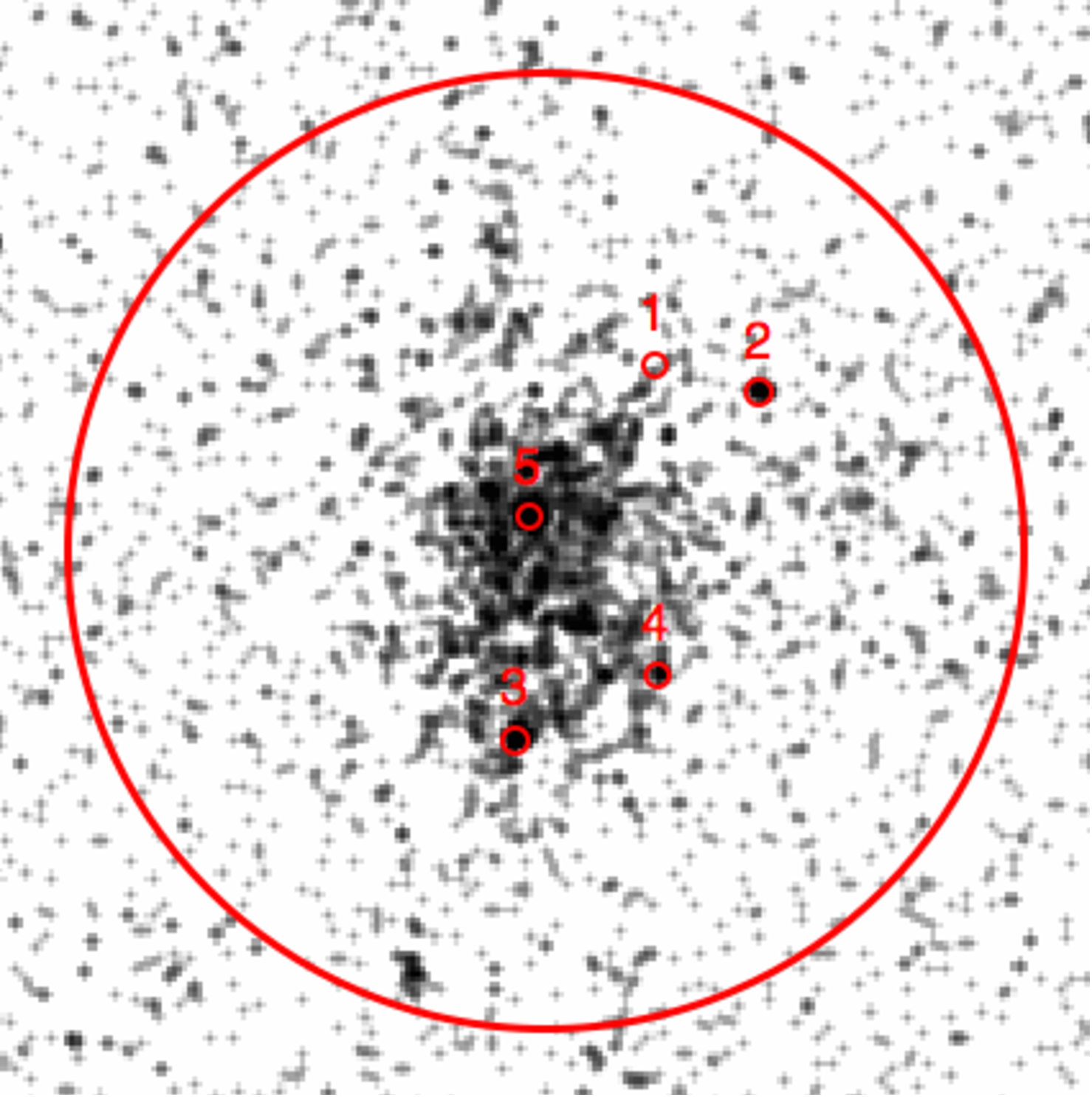}
\includegraphics[width=10cm,angle=0]{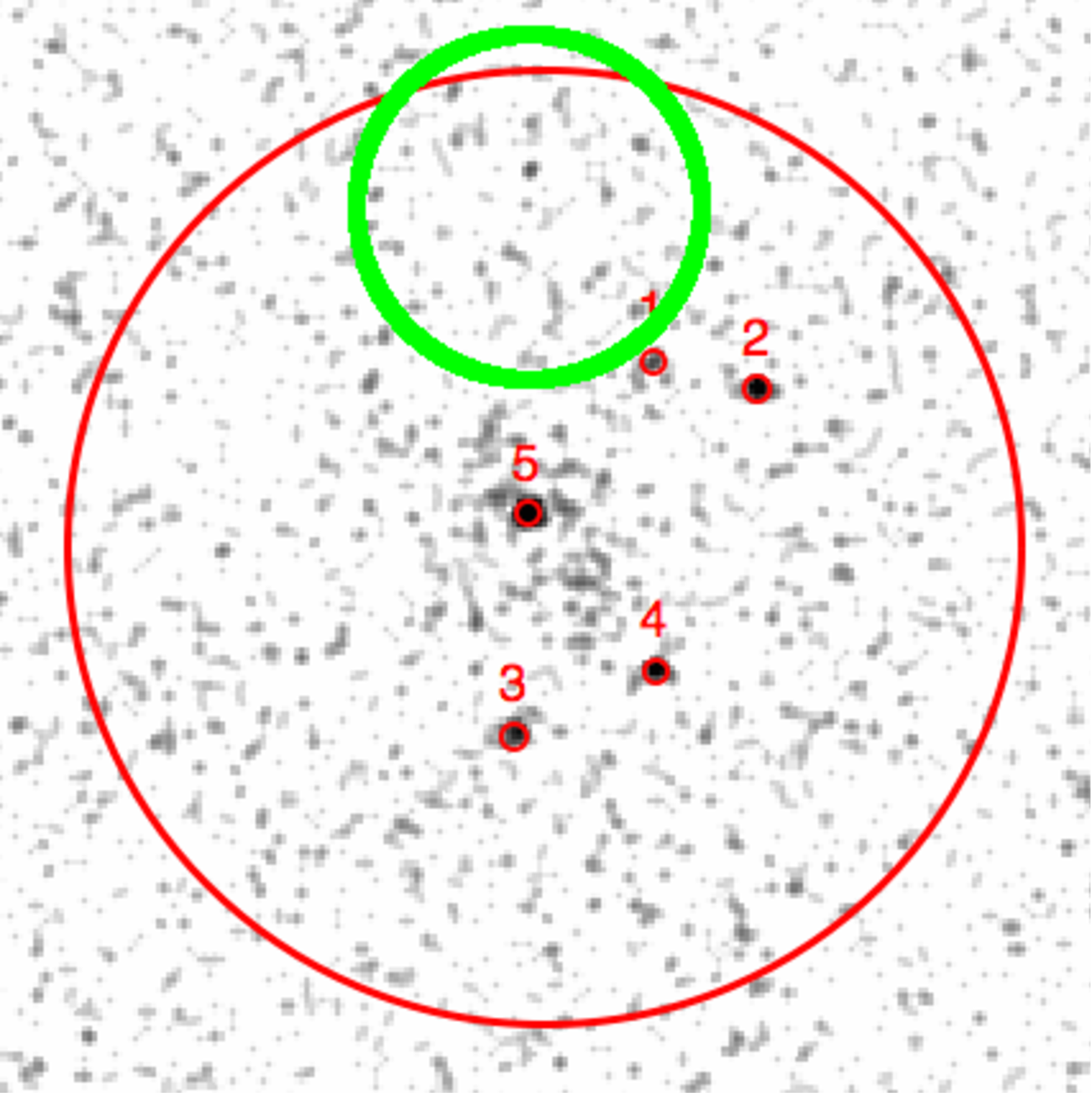} 
\caption{\label{closeup}Chandra image of XDCP0044 in the soft 0.5-2 keV band (left
  panel) and in the hard 2-7 keV band (right panel).  The image has not been
  rebinned (1 pixel corresponds to $0.492\arcsec$).  The large red
  circle shows the extraction region used for the spectral analysis,
  corresponding to $R_{ext}=44"=375 $ kpc at $z=1.58$.  The five small circles
  show the unresolved sources identified in the soft or in
  the hard band images.  The green circle in the hard-band image shows the uncertainty in the position of the radio source NVSS 004405-203326.  Images are $1.7\times 1.7$ arcmin across.}
\end{figure}

In Figure \ref{closeup} we show a close-up of the XDCP0044 X-ray image in
the soft (left panel) and hard (right panel) bands.  By running the {\tt ciao} detection algorithm 
{\tt wavdetect} we identify five unresolved sources overlapping with the
ICM emission.  We carefully check by eye that the detection algorithm 
kept well separated the unresolved and the surrounding extended emission, to be able to remove only the
unresolved source contribution and eventually extract the ICM emission free from AGN contamination.  We
measure the aperture photometry for the five unresolved sources within a radius of $2.5\arcsec$, which 
is expected to include about 90\% of the flux, exploiting the exquisite angular resolution of {\sl Chandra} close to
the aimpoint.   The background is estimated locally in an annulus around each source (outer radius $10"$, inner radius $5"$) 
in order to accurately take into account the surrounding extended emission.
The approximate soft and hard band fluxes are computed after correcting for vignetting and assuming the typical
conversion factors in the soft and hard band corresponding to a power
law emission with photon index $\Gamma = 1.4$.   Considering the effective area of each single exposure at the position of the cluster,  
and assuming a Galactic absorption column density of $N_{H}= 1.91 \times 10
^{20}$cm$^{-2}$ (see next Section) we obtain $C_{soft} = 4.93 \, \times 10^{-12}$ cgs/cts/s and $C_{hard} = 2.64\, \times 10^{-11}$
cgs/cts/s.   We find that the total soft band flux contributed by point sources is about 12\% of the diffuse emission, in line with what is found
in non X-ray selected clusters \citep[see, e.g.,][]{2008Bignamini}.  Incidentally, this confirms that the presence of X-ray emitting AGN 
within the extended emission of distant clusters does not hamper one from identifying them even with moderate angular resolution.
The results are shown in Table \ref{Table1}.  Note that while the net detected counts in the hard band
refer to the 2-7 keV energy range, we compute the energy fluxes in the 2-10 keV band, for a better comparison with the literature,
despite our unresolved sources show very little signal above 7 keV.  At least two of them (\# 3 and \# 5) are cluster members, and one (\# 4) is a foreground AGN.  The two unidentified
X-ray sources are likely to be foreground AGN due to their hard X-ray emission.  A further X-ray source is just on the edge of the extraction regions, 
and it is identified with a foreground galaxy.  There are other unresolved X-ray sources within the virial radius of the cluster, which potentially can be 
cluster members. A more detailed discussion of the cluster galaxy population, including their X-ray properties, will be presented in forthcoming papers.

\begin{table}
 \centering
\caption{Aperture photometry of the five unresolved sources identified within the X-ray extended emission of XDCP0044.
The photometry corresponds to the net number of counts detected within an aperture radius of $2.5"$ in the soft (0.5-2 keV) and hard band (2-7 keV) images after
background subtraction.  Energy fluxes are obtained in the soft and hard (2-10 keV) bands assuming a power law spectra with photon index $\Gamma = 1.4$ and 
taking into account Galactic absorption. Sources \# 3 and \# 5 are confirmed cluster members, while source \# 4 is a foreground AGN.}
\label{Table1}

\begin{tabular}{@{}cccccc@{}}\hline
Source & Soft counts      & Hard counts &     Soft flux  & Hard flux   & redshift \\
            & (0.5-2 keV)     &  (2-7 keV)    &  (0.5-2 keV)   &     (2-10 keV)   & \\
\hline  
  \# 1       &   $1.8\pm 2.0$   &    $7.8 \pm 3.0$  &  -	&	($5.6\pm 2.2) \, \times  10^{-16}$	 &  - \\ 
 \# 2       &   $15.2\pm 4.1$   &    $29.0 \pm 5.5$   & ($2.0\pm 0.5) \, \times 10^{-16}$ & 	($2.1\pm 0.4) \,\times 10^{-15}$&   - \\ 
 \# 3       &   $40.0\pm 7.0$   &    $20.0 \pm 5.0$  &  $ (5.4 \pm 0.9) \, \times 10^{-16}$   & $(1.4\pm 0.4)\,\times 10^{-15}$&  $1.5703$ \\ 
 \# 4       &   $8.4\pm 3.7$   &    $25.8 \pm 5.3$   & $(1.1\pm 0.5) \,\times 10^{-16}$ 	& $(1.9\pm 0.4) \,\times 10^{-15}$ &  $0.5922$ \\ 
\# 5       &   $122.0\pm 12.0$   &    $117.0 \pm 11.0$  & $(1.6\pm 0.15)\,\times 10^{-15}	$ & $(8.4\pm 0.8)\,\times 10^{-15}$ &   $1.5785$ \\ 
\hline
\end{tabular}
\end{table}

The nature of the X-ray extended emission after the removal of the unresolved source contribution
can be safely assumed to be entirely due to the thermal bremsstrahlung from the ICM.  We can exclude a significant contribution from Inverse-Compton 
emission associated to a population of relativistic electrons, since such a component would result in radio emission visible with
NVSS data.   The NVSS radio image shows no radio emission in the direct vicinity of the X-ray center.  However, the NVSS catalog reports a weak radio
source, NVSS 004405-203326, at a distance of $33\arcsec$ from the cluster center, with an estimated flux of $3.2 \pm 0.6 $ mJy at 1.4 GHz and an uncertainty in the position of about $15\arcsec$
\citep{1998Condon}.  The position of the radio source is shown as a green circle in Figure \ref{closeup}, right panel, and it is clearly unrelated to the X-ray emission.

We perform simple aperture photometry of the extended emission after removing the detected point sources.  First, we compute the centroid 
of the X-ray emission, by searching the position where the aperture photometry within a fixed radius
of 100 kpc returns the highest S/N in the 0.5-7 keV band.  We find that
the center of the X-ray emission is in RA$_{X}=00$h$44$m$05.27$s, DEC$_{X}=-20^\circ 33^\prime 59.4^{\prime\prime}$.  
We also notice that given the flat and irregular surface brightness distribution, the X-ray centroid can vary by $\sim 3\arcsec$ when the aperture radius 
ranges between 20 and 200 kpc.  Therefore, despite the lack of a well defined
peak in the surface brightness distribution, the X-ray centroid is relatively stable and in excellent agreement with the XMM position.

\begin{figure}
\begin{center}
\includegraphics[width=14cm,angle=0]{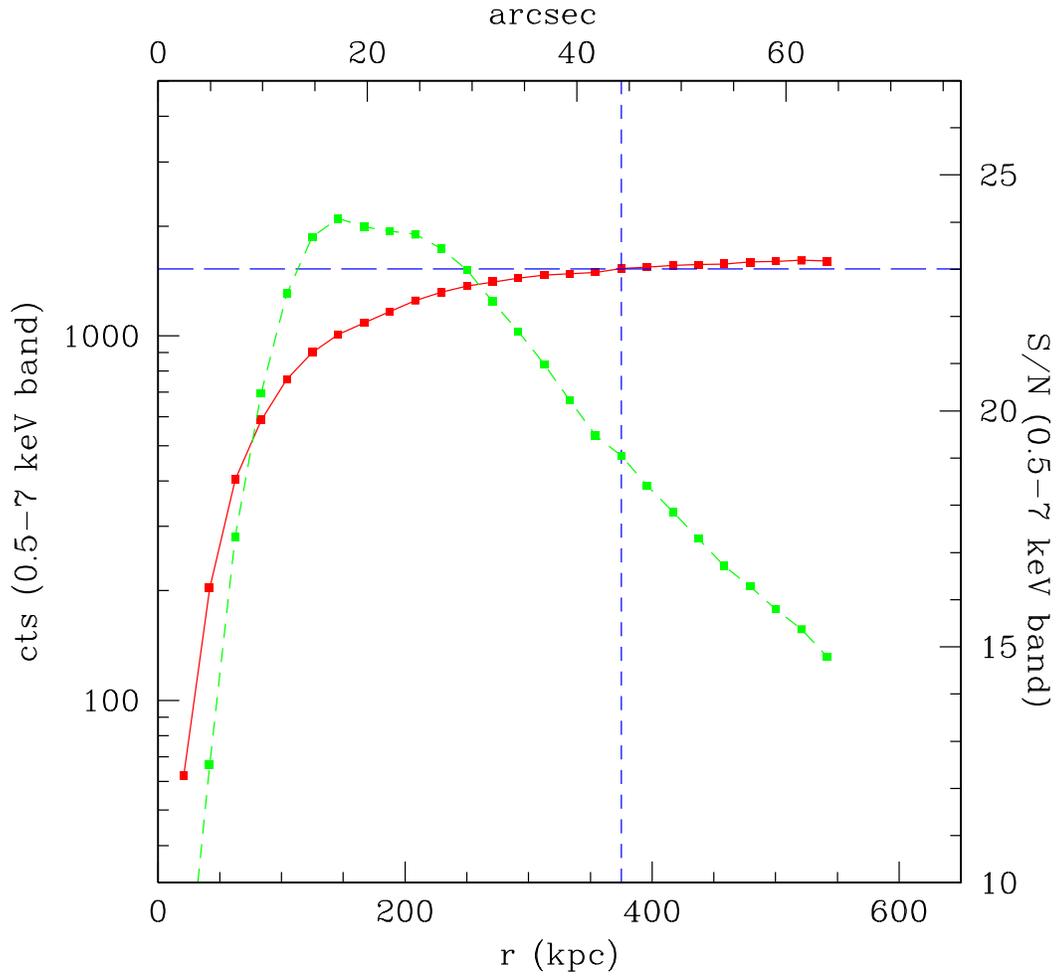} 
  \end{center}
\caption{  \label{photometry}The solid red line shows the aperture photometry in the total (0.5-7 keV) band for
  XDCP0044 as a function of the physical projected radius. The dashed
  vertical line is the extraction radius used for spectral analysis, corresponding to $R_{ext}=375$ kpc. 
  The green dashed line is the measured S/N within a given radius. The horizontal line is the photometry 
  measured within $R_{ext}$  corresponding to $1500 \pm 80$ net counts in the 0.5-7 keV band. }
\end{figure}

The background is estimated from an annulus centered on the cluster and
distant from the cluster emission.  The outer radius of the background annulus is $137"$, while the inner radius is $79"$. 
Clearly all the identified unresolved sources in the background region are removed before extracting the spectrum.
 We measure a background of $3.8 \times 10^{-7}$ cts/s/arcsec$^2$ in the
0.5-2 keV band and $6.2 \times 10^{-7}$ cts/s/arcsec$^2$ in the 2-7 keV band.
The total net counts detected in the 0.5-7 keV band as a function of the
physical projected radius are shown in Figure \ref{photometry}.  We
find $ 1500 \pm 80 $ net counts within 375 kpc.  The extended emission lost because of the removal
of circular regions of radius $2.5"$ around unresolved sources has been estimated to be $\sim 20$ net counts in the
same band, about $1.3$\%, well below the 1 $\sigma$ poissonian uncertainty
(about $\sim 5$\%). 

We also investigate whether our choice of the background affects the photometry, and, eventually, the spectral
analysis.  We repeat the photometry with a synthetic background obtained by processing the Chandra ACIS blank sky files in the same
way we processed our data \footnote{http://cxc.harvard.edu/ciao/threads/acisbackground/}.  
We reproject the background events according to each separate obsid, and normalize the background for each obsid by choosing source-free
regions in the actual data in several different regions of the detector and requiring the same 0.5-7 keV
average count rate.  We finally obtain a synthetic background image we can use to
subtract to the data image.  With this procedure we find $1494\pm 84$ net counts, consistent well within 1 $\sigma$
with the photometry based on the local background.  

\subsection{Spectral analysis: temperature and iron abundance}

We perform a spectral analysis of the ACIS-S data with {\tt Xspec}
v12.8.1 \cite{arnaud96}.  The adopted spectral model is a
single-temperature {\tt mekal} model \citep{kaa92,lie95}, using as a
reference the solar abundance of \citet{asp05}.  The local absorption
is fixed to the Galactic neutral hydrogen column density measured at
the cluster position and equal to $N_{H}= 1.91 \times 10
^{20}$cm$^{-2}$ taken from the LAB Survey of Galactic HI
\citep{2005LAB}.  The fits are performed over the energy range
$0.5-7.0$ keV.  We used Cash statistics applied to the source plus
background, which is preferable for low S/N spectra \citep{nou89}.

\begin{figure}
\begin{center}
\includegraphics[width=10.0cm,angle=-90]{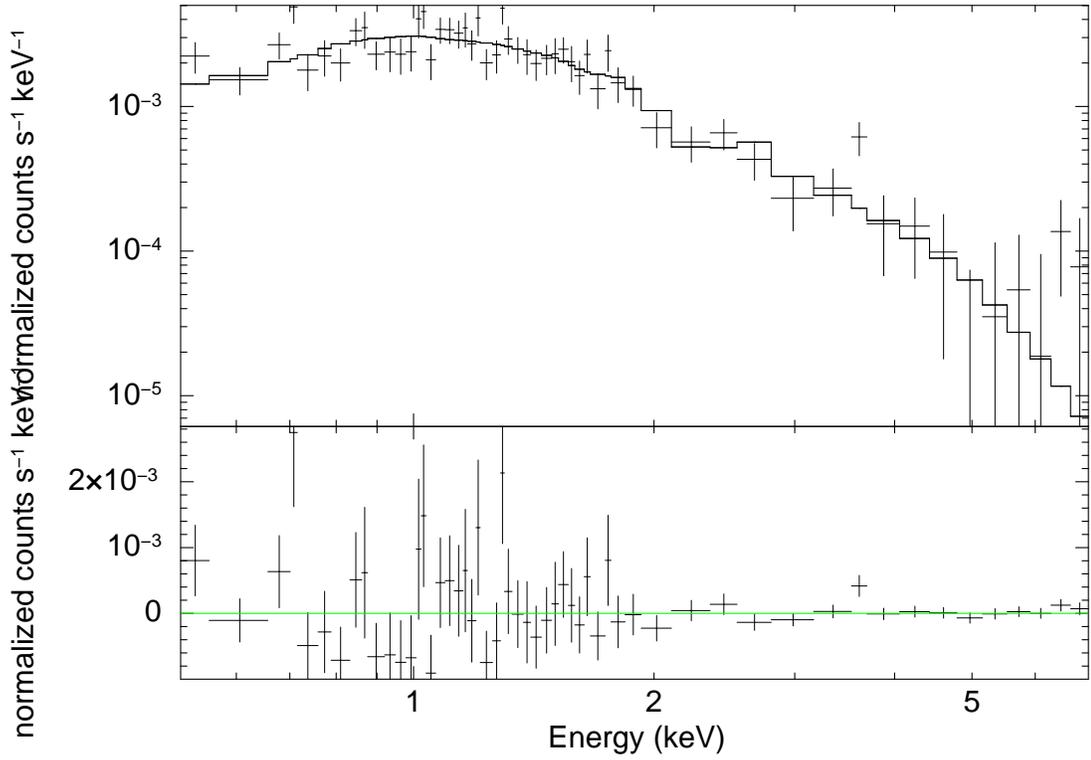}
  \end{center}
\caption{F\label{spectrum}olded \textit{Chandra} ACIS-S spectrum of XDCP0044 within $44"$ (375 kpc) with its
  best-fit {\tt mekal} model (solid line).  The presence of the iron $K_\alpha$ lines
  complex, expected at 2.6 keV in the observed reference frame, has a statistical significance below 2 $\sigma$.  The lower panel show the residuals.  The spectrum has been 
  binned with a minimum of 25 counts per bin only  for plotting purpose.}
\end{figure}

We extract the cluster emission from a circle of radius $R_{ext}=44"$,
corresponding to 375 kpc at $z=1.58$.  This radius is chosen in order
to encompass the maximum signal ($\sim 1500$ net counts in the 0.5-7 keV band, see Figure \ref{photometry}).  
Beyond the radius $R_{ext}$, the residual signal is consistent with zero within the statistical
1 $\sigma$ error on photometry.  The spectrum of the total emission within the extraction radius, after the removal of the five 
unresolved sources discussed in Section 2.3, is fitted with the single-temperature {\tt mekal} model.  First, we leave the redshift parameter free to vary, but we are not
able to obtain a reliable measurement of the X-ray redshift $z_X$.  For \textit{Chandra} observations, it was estimated that about 1000 net counts are needed
to measure $z_X$ at a $3\sigma$ confidence level, and even more are
needed in the case of hot clusters, as shown in \citet{2011Yu} for medium exposures ($\sim 100 $ ks) .  In
the case of XDCP0044, the large exposure time also implies a relatively larger background and lower S/N with respect to the typical case explored in \citet{2011Yu}, 
and this prevents us from retrieving the correct redshift directly from the X-ray spectral
analysis, even with  $\sim 1500$ net counts.  Then, we set the redshift to $z=1.58$ and find that the 
best-fit global temperature is $kT= 6.7_{-0.9}^{+1.3}$ keV.  The measured iron abundance, in units of \citet{asp05}, is $Z_{Fe} =
0.41_{-0.26}^{+0.29}Z_{Fe_\odot}$.  The cluster unabsorbed soft-band flux within a
circular region of $44\arcsec$ radius is $S_{0.5-2.0 keV}=(1.66 \pm
0.09) \times 10^{-14}$ erg s$^{-1}$ cm$^{-2}$ (in excellent agreement with the soft flux estimated from the XMM-Newton discovery data), and the hard flux
$S_{2-10 keV}=(1.41 \pm 0.07) \times 10^{-14}$.  At a redshift of $z=1.58$ this correspond to a rest frame soft band luminosity of $L_S = (1.89
\pm 0.11) \, 10^{44}$ erg/s.  The X-ray bolometric luminosity is $L_{bol} = (6.8 \pm 0.4) \, 10^{44}$ erg/s. 
In Figure \ref{spectrum} we show the spectrum of XDCP0044 within a radius of 375 kpc along with the best-fit model.

We also perform the spectral analysis in the region with the maximum S/N, 
corresponding to a radius of $17"$ ($145$ kpc; see Figure \ref{photometry}).  This region
includes about 2/3 of the signal we found in the $R_{ext}=44"$ region.  
We obtain $kT=5.9_{-0.5}^{+0.8}$ and $Z_{Fe} = 0.28_{-0.18}^{+0.20}Z_{Fe_\odot}$.  In this case, if the redshift
is left free, we are able to recover a best-fit redshift of $z_X =  1.59_{-0.06}^{+0.08}$.  Despite this
value is in perfect agreement with the optical redshift, the significance level of the X-ray redshift is still below
$2 \sigma$.  


We highlight that, in both cases, the measured iron abundance is consistent with zero within 2 $\sigma$.  This marginal detection does
not allow us to draw any conclusion on the possible evolution of the iron abundance  of the ICM in XDCP0044 with respect to local clusters.
However, this finding 
is consistent  with the mild, negative evolution of the iron abundance of a factor 1.5-2 between 
$z=0$ and $z\sim 1$  found in previous works
\citep{2007Balestra,2008Maughan,2009Anderson,2012Baldi}.  We conclude that, despite our expectations, we are not able to use XDCP0044 to put significant constraints on the
iron enrichment time scale in the ICM.  This study is currently limited to the brightest clusters at $z \sim 1$ \citep[see][]{2014Degrandi}.  
A systematic investigation of the iron abundance evolution in very high-z clusters
requires higher sensitivity, and it will be an important science case for future X-ray facilities such as Athena \citep{2012Barcons,2013Nandra} and SMART-X\footnote{http://smart-x.cfa.harvard.edu/}.

\begin{figure}
\includegraphics[width=13.0cm]{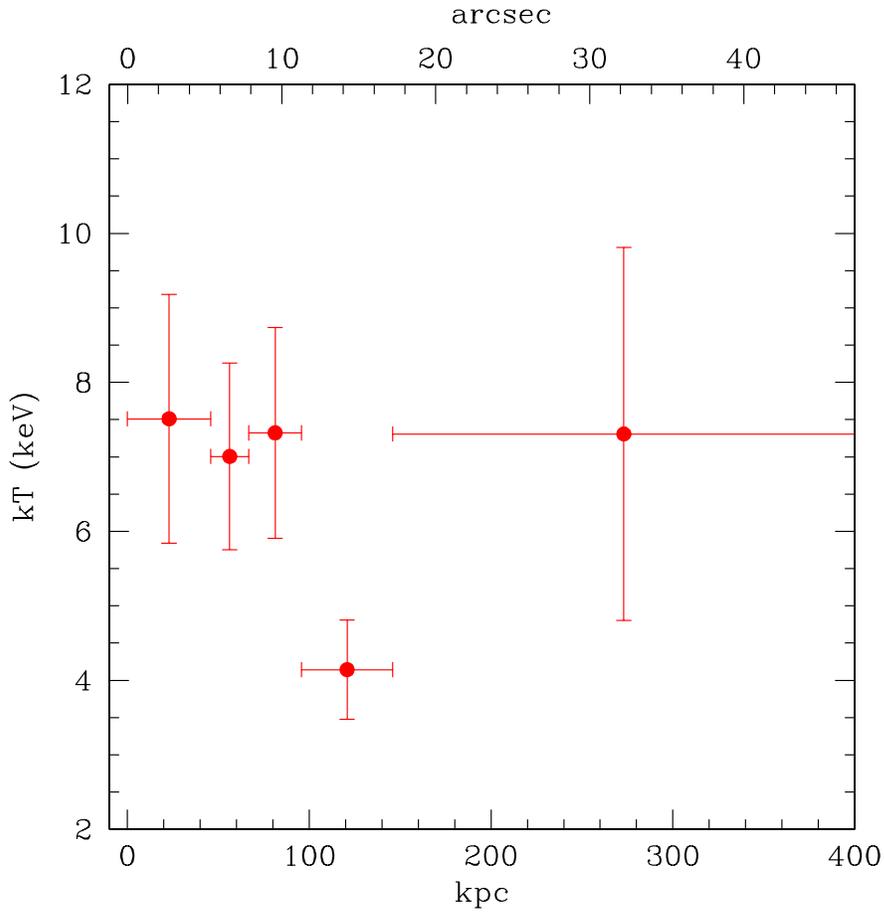}
\caption{\label{Tprofile}Projected temperature profile from the spatially resolved spectral analysis of XDCP0044.  Each bin includes 
about 300 net counts in the 0.5-7 keV band.}
\end{figure}

\begin{figure}
\begin{center}
\includegraphics[width=14.0cm]{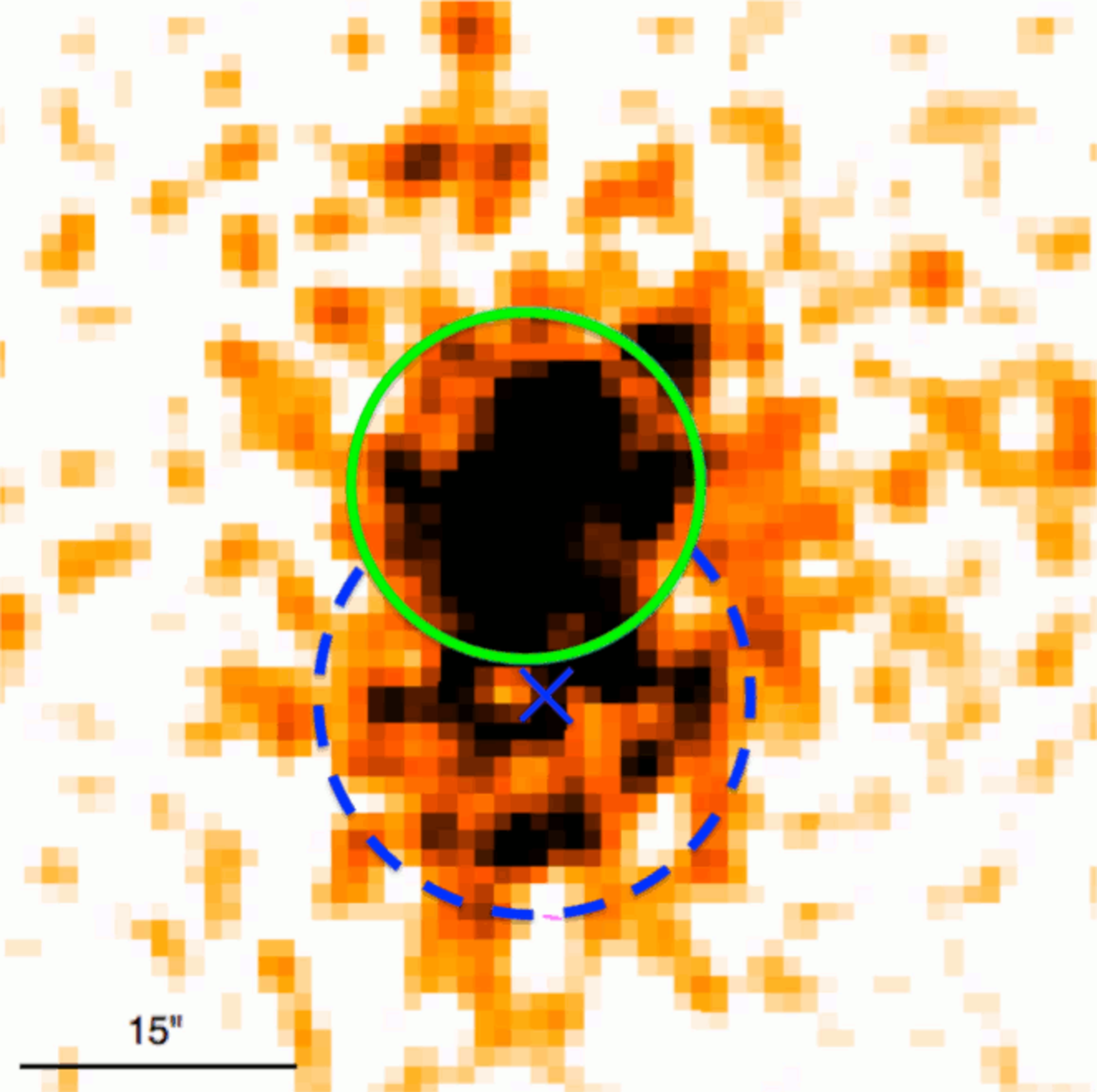}
\end{center}
\caption{\label{clumps}North and South clumps of XDCP0044 identified by visual inspection.  
The extraction region of the North clump is given by the entire circular region (green solid line), while for the South
clump is given by a circular region after excluding the overlap with the North region (dashed blue circle).  }
\end{figure}

We also attempt a spatially resolved spectral analysis  dividing the ICM emission in five annular
bins with about 300 net counts each.  We tentatively find a hint of a decreasing
temperature beyond 100 kpc, as shown in Figure \ref{Tprofile}, which
also explains the 1 $\sigma$ difference between the two extraction
regions at $44"$ and $17"$.  We perform a preliminary study to investigate whether this decrease may
be associated to a difference in temperature in two different regions
of the clusters.  To do that, we identify, by visual inspection, two
clumps (North and South) as shown in Figure \ref{clumps} where the extraction regions are drawn
manually.   The two clumps are identified based on a significant difference in
the average surface brightness, which in the North clump is measured to be about 1.8 times 
higher than in the South clump at more than 4 $\sigma$ confidence level.   The reliability of the existence
of two physically different regions is also suggested by the observed discontinuity in the surface
brightness profile at a distance from the center corresponding to the
dividing line of the North and South clumps (see \S 2.5).   The North
clump is identified by a circular region centered in $00$h$44$m$05.4$s,
$-20^\circ33^\prime 56.2^{\prime\prime}$ with a radius of $10\arcsec$.  The South clump is
identified by a circular region centered in $00$h$44$m$05.3$s,
$-20^\circ 34^\prime 07.6^{\prime\prime}$ and a radius of $12\arcsec$, with the exclusion of the
part overlapping with the North clump.  The distance between the center of the two clumps
amounts to $11.6"$, corresponding to $\sim 100 $ kpc.  The soft-band fluxes in the
North and South clumps are $5.4\times 10^{-15}$ erg cm$^{-2}$ s$^{-1}$
and $3.5 \times 10^{-15}$ erg cm$^{-2}$ s$^{-1}$, respectively.  For
the North clump we obtain $ kT= 7.1_{ -1.0}^{+1.3}$ keV and for the
South clump $kT= 5.5_{-1.0}^{+1.2 }$ keV.  The difference is comparable to the 1 $\sigma$ uncertainty and therefore we are not able to
confirm a temperature gradient between the two regions.  However, this
finding, coupled to the irregular appearance of the surface brightness
of XDCP0044, points towards a complex dynamical state of the cluster.
Another hint is an extension of diffuse emission for approximately
15 kpc above the North clump (see Figure \ref{clumps}).  However, in
this case the emission is too faint to attempt a spectral analysis.  Finally, a wide Western extension is also visible in the contours in Figure 1.
Despite these hints for a complex, not spherically symmetric morphology, in the rest of the Paper we will
assume spherical symmetry to measure the gas and total masses as a function of radius, a choice which is tested in \S 2.5.  Further discussion on
the presence of substructures in the ICM is postponed to Section \S 4.  For
the same reason, we will rely on the spectral analysis of the emission within
$R_{ext}=44"$, which includes the maximum useful signal.

Finally, we investigate whether our spectral analysis is robust against different choices of the background.  We repeat the spectral
analysis using the synthetic background extracted from the same background region and from the same source region of our data.  
We remind that the synthetic background is obtained summing the contribution of the synthetic background for each Obsid, where the count rate in the 0.5-7 keV band is normalized to that
measured in the data within the same region. Using the synthetic background spectrum extracted from the background region, we find $kT= 7.8_{-1.3}^{+1.5} $ keV and $Z_{Fe}=
0.19_{-0.19}^{+0.26} Z_{Fe_\odot}$.  If, instead, we use the synthetic background extracted from the source region, we find  $kT=
8.2_{-1.5}^{+2.0}$ keV, $Z_{Fe}= 0.40_{-0.33}^{+0.35} Z_{Fe_\odot}$.
In general, the use of a synthetic background normalized to 0.5-7 keV count rate in the data, provides slightly harder
spectra, and therefore higher temperatures.  However, the difference is always consistent within 1 $\sigma$ with our analysis based on the
local background.  
In the rest of the Paper, we will rely only on the spectral analysis obtained with the local background.

\subsection{Surface brightness profile and concentration}

\begin{figure}
\includegraphics[width=14.0cm,angle=0]{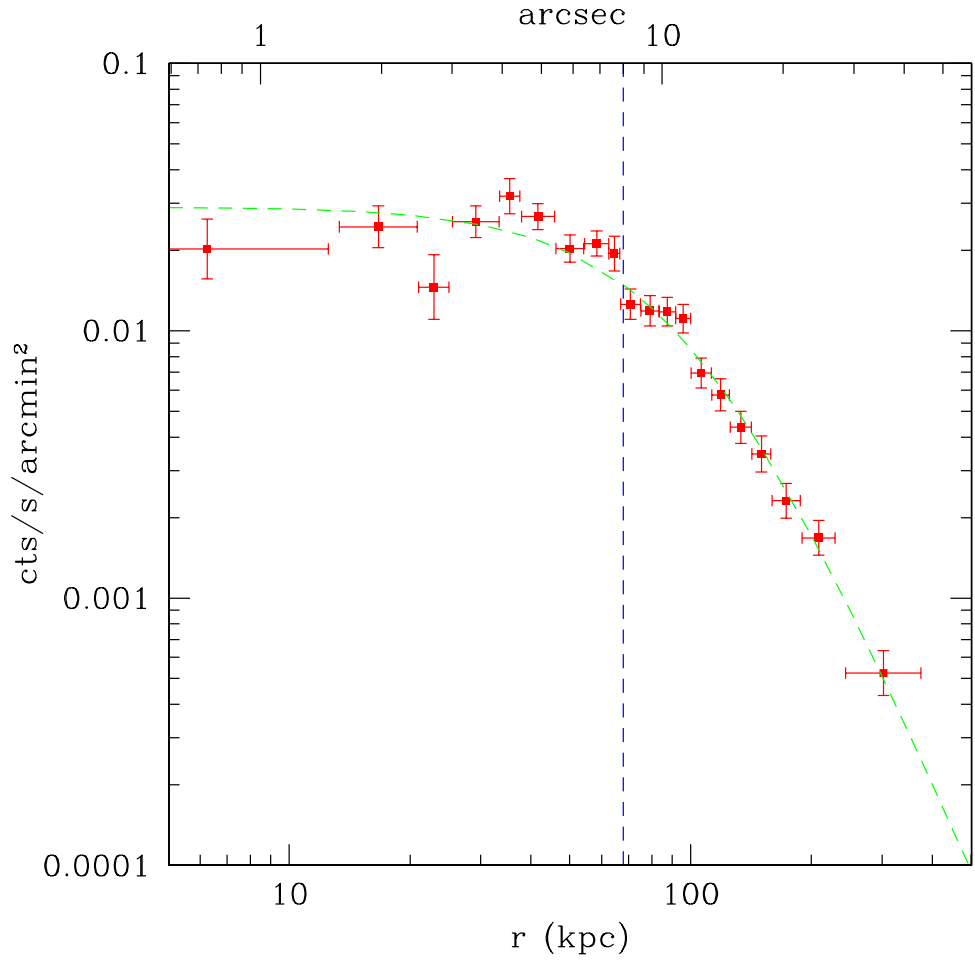}
\caption{\label{SB}Background-subtracted surface brightness profile in the soft (0.5-2 keV) band (points) and best-fit $\beta$ model (green dashed line) for XDCP0044.  
Error bars correspond to 1 $\sigma$ uncertainty.  The vertical dashed line corresponds to the distance of the border of the Northern clump from the X-ray centroid}.
\end{figure}

\begin{figure}
\includegraphics[width=14.0cm,angle=0]{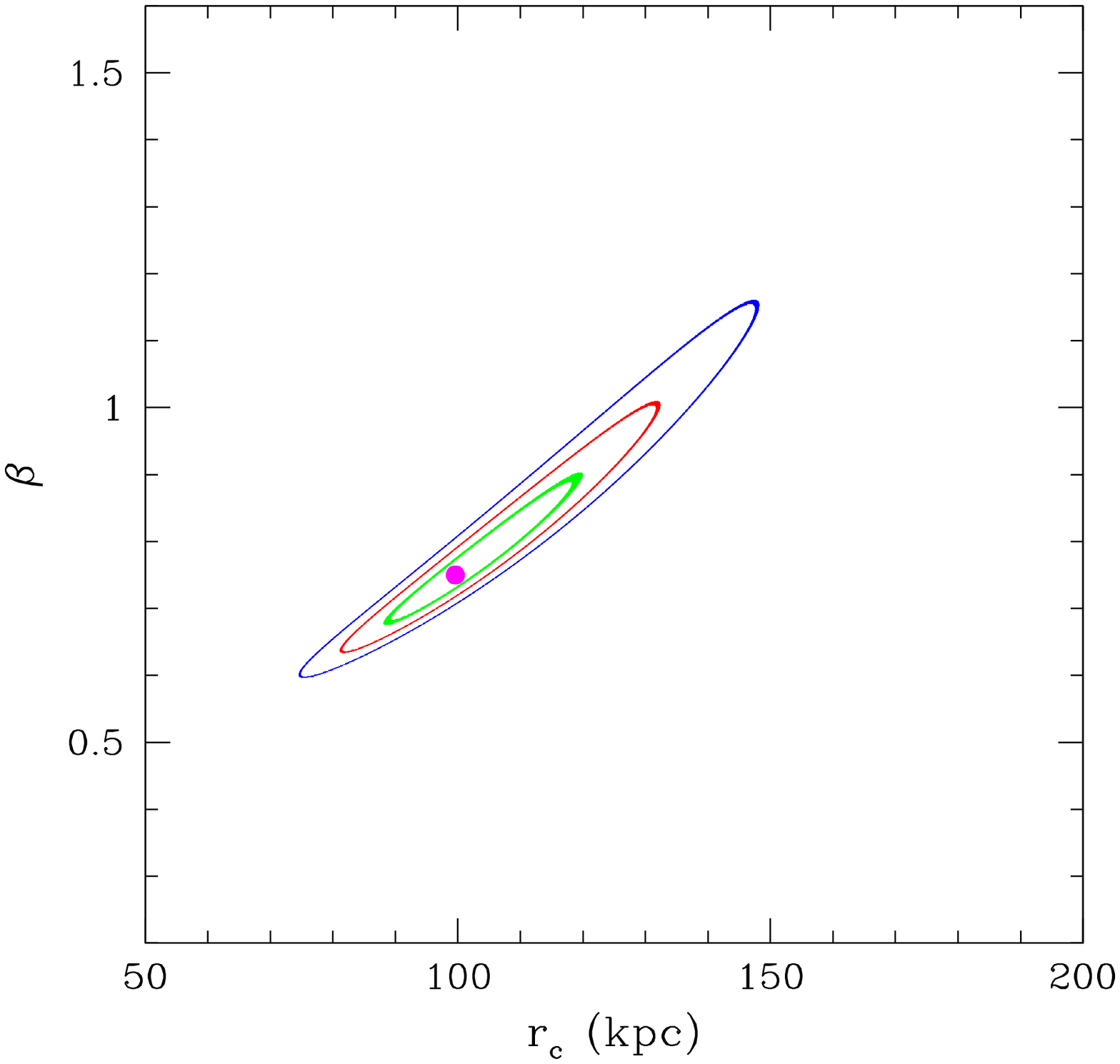}
\caption{\label{SBpar} Green, red and blue contours show the 1, 2 and 3 $\sigma$ confidence levels, respectively, for two
  relevant parameters $\beta$ and $r_c$ of the surface brightness profile. }
\end{figure}

We compute the azymuthally averaged surface brightness profile from the exposure-corrected ACIS-S image in the soft band out to $R_{ext} =
44"$ and centered in RA$_{X}=00$h$44$m$05.27$s, DEC$_{X}=-20^\circ 33^\prime 59.4^{\prime\prime}$.  The profile is well described by a single $\beta$ model, $S(r)
= S_{0} (1+(r/r_{c})^2) ^{-3\beta+0.5} + bkg$ \citep{cav76}.  The azymuthally averaged surface brightness profile and its best-fit model
are shown in Figure \ref{SB}.  We note that the surface brightness distribution in the core region is noisy, and this does not depend on the choice of the X-ray center.
Incidentally, we remark that the first bin is a circle with a radius of 3 pixels, corresponding to $1.5"$, which is the smallest region which can 
be resolved by {\sl Chandra} at the aimpoint. 
As we already mentioned, this suggests a lack of a well developed core.  The best-fit parameters are $\beta=0.75\pm 0.04$ and
core radius $r_{c} = 100\pm 10$ kpc for a reduced $\tilde \chi^2 =
1.37$.  The surface brightness profile is well rendered by the
$\beta$-model and it allows us to derive accurate gas mass
measurements despite the degeneracy between the parameters $\beta$ and
$r_c$ (see Figure \ref{SBpar}).  However, the extrapolation of the
surface brightness profile at radii larger than $R_{ext}$ is
uncertain, and further assumptions are needed to estimate the total
mass beyond this radius.  We also note that the azymuthally averaged
surface brightness profile shows some irregularities.
However, we detect a discontinuity in the surface brightness profile at
a radius of $\sim 70$ kpc, at a $2 \sigma$ confidence level.  
We argue that this feature may be associated to the presence of the two clumps discussed in Section \S 2.4. 
In fact, the center of the Northern clump is only $3.3"$ from the center of the overall X-ray emission, used to compute the 
azimuthally averaged surface brightness. Since the radius of the Northern clump is $10"$, the distance where the surface brightness starts to be sampled beyond the 
Northern clump is roughly $70$ kpc.  Therefore, the jump in the azimuthally averaged surface brightness profile exactly at 70 kpc can be ascribed to the 
sudden change in surface brightness outside the Northern clump.  This finding, despite with low statistical significance, 
provides another hint  of a physical difference between the two regions, possibly associated to an ongoing or recent merger, or to a very recent virialization for the
bulk of the cluster.

We also attempt a double  $\beta$ model fit, and we find that an  improvement in the $\chi^2$ of $\Delta \chi^2 \sim 2.4 $, while the
reduced $\tilde \chi^2 = 1.50$ is higher than in the single $\beta$ model fit, showing no significant improvements. 
Furthermore, the measurement of the source profile in sectors is not feasible given the 
low signal.   Therefore, we conclude that the assumption of spherical symmetry and the use of a single  $\beta$ model, are fully acceptable for our data, 
while more complex models would introduce unnecessary degrees of freedom.

From the surface brightness profile we can compute the correction factor to be applied to the measured luminosity within $R_{ext}$ to
obtain the total luminosity within the virial radius.  This factor is 1.11-1.13 for $R_{vir}=800-1000 $
kpc.  Applying this factor to the luminosity measured within $R_{ext}$ most likely provides an upper limit to the total luminosity, given that
surface brightness profiles are generally observed to steepen in the outer regions with respect to the extrapolation of the $\beta$-fit model
\citep{ettori09b}.

As already mentioned, the surface brightness profile within 100 kpc does not support the presence of a cool core.  To quantify the core strength,
we compute the value of $c_{SB}$, a concentration parameter defined in \citet{santos08} as the ratio of the surface brightness within $40$
kpc and $400$ kpc: $c_{SB}=SB\,(<40$ kpc$)\,/\,SB\,(<400$kpc$)$.  A cluster is expected to host a cool core if $c_{SB} > 0.075$.  This simple
phenomenological parameter has proven to be a robust cool core estimator, particularly in low S/N data typical of distant clusters.
We measured $c_{SB}=0.12 \pm 0.02$, which is 2 $\sigma$ above the minimum $c_{SB}$ expected for a cool-core cluster.  However, in the
case of XDCP0044 we notice that the surface brightness is well sampled only within $ \sim 100$ kpc,  much less than the reference 
radius of $400$ kpc, and this may introduce some errors in the measured $c_{SB}$ value.  If we compare the best-fit profile of the electron density of XDCP0044 with that of the
high-z cool-core cluster WARP1415 \citep{2012Santos}, we see a remarkable difference, despite the similar measured $c_{SB}$.
Actually, the $n_e(r)$ profile of XDCP0044 is very similar to that of CXO1415 \citet{2013Tozzi}, a comparable massive cluster at $z\sim
1.5$. If we use the temperature and electron density measured within 100 kpc, we find e central entropy of $94^{+19}_{-14}$ keV
cm$^2$, corresponding to an average cooling time of $4.0^{+0.9}_{-0.6}$ Gyr for an iron abundance  $\sim 0.4 Z_{Fe\odot}$.  Despite
these are average values within 100 kpc, the rather flat density profile and the lack of any hint of a temperature drop within 100 kpc suggest that
these values are representative of the cluster inner region, and clearly classify XDCP0044 as a non-cool core cluster.  
Since cool cores are already established at $z\sim 1$ \citep[][]{2012Santos}, we may begin to identify a population of massive clusters at an epoch before the cool
core appearance.  Clearly this can be tested only with much larger statistics and high quality data, a goal which is challenging given
the limited discovery space of present-day X-ray observatories.
 
\section{Gas mass and total mass}

The baryonic mass can be directly computed once the three-dimensional
electron density profile is known.  For a simple $\beta$ model, this is
given by $n_e(r) = n_{e0}[ (1+(r/r_{c})^2) ^{-3\beta/2}]$.  To measure
$n_{e0}$, we use the relation between the normalization of the X-ray
spectrum and the electron and proton density in the ICM for the {\tt
mekal} model:

\begin{equation} 
Norm = {{10^{−14}}\over { 4\pi \, D^2_a\, (1 + z)^2}}\times \int
n_en_HdV \, ,
\end{equation}

\noindent
where $D_a$ is the angular diameter distance to the source (cm), $n_e$ and $n_H$ (cm$^{-3}$) are the electron and hydrogen densities,
respectively, and the volume integral is performed over the projected region used for the spectral fit.  For XDCP0044, this
gives a central electron density of $n_{e0} = (2.15 \pm 0.13) \times 10^{-2} $ cm$^{-3}$.  
The physical three-dimensional electron density distribution can be obtained simply by deprojecting the best fit $\beta$-model.

\begin{figure}
\includegraphics[width=14.0cm]{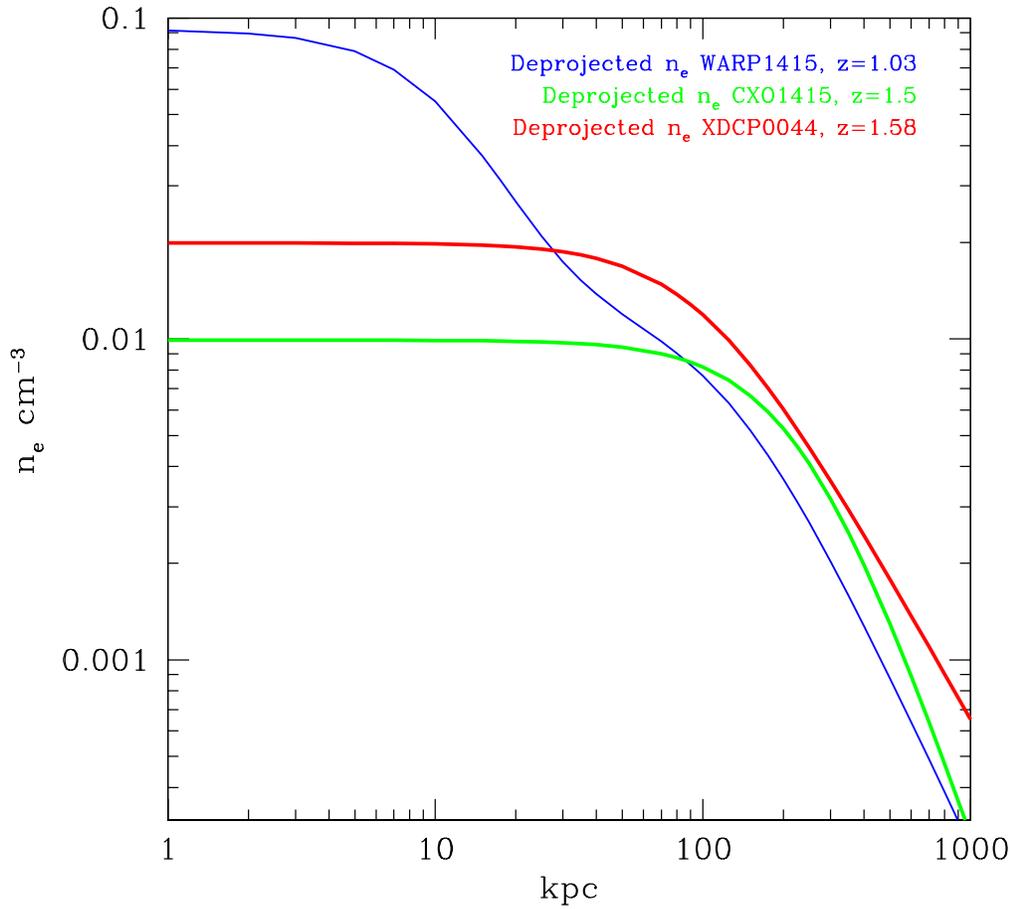}
\caption{\label{ne_profile}Best-fit to the deprojected electron density profile of XDCP0044 (red line) compared with those of CXO1415 \citep[green line, ][]{2013Tozzi} 
and W1415 \citep[blue line, ][]{2012Santos}.}
\end{figure}

In Figure \ref{ne_profile} we show the analytical best-fit of the electron density $n_e(r)$ compared with that found in  W1415 and CXO1415.  
The former is a $z\sim 1$ strong cool-core cluster with a lower temperature \citep{2012Santos}, showing a density profile strikingly different 
from that of XDCP0044.  On the other hand, the density profile of XDCP0044 is quite similar to that of CXO1415 is a comparable, $z\sim 1.5$ clusters with a lower luminosity and a
comparable global temperature $kT = 5.8_{-1.0}^{+1.2}$ keV \citep{2013Tozzi}.
The difference in luminosity (the bolometric luminosity of XDCP0044 is more than three times that of CXO1415) can be partially explained with
the difference in the temperatures of approximately 1.3 keV.  Both clusters appears to be a factor
of two less luminous than expected on the basis of the empirical, redshift-dependent $L-T$ relation of \citet{2009Vikhlinin}.
However, recent data of distant clusters support a weaker luminosity evolution
with respect to what is found in \citet{2009Vikhlinin} from a much smaller redshift range \citep[see][]{2011Reichert,2012Boehringer}.
A factor of two difference at this high redshift is consistent with this scenario and in line with preheating models.

The spectral analysis and the surface brightness analysis allow us to
have a direct measure of the total ICM mass within the extraction
radius $R_{ext} = 375$ kpc, which yields  $M_{ICM}=(1.48\pm
0.2) \times 10^{13} M_\odot$.  On the other hand, the total cluster
mass can be estimated under the assumptions of hydrostatic equilibrium
and spherical symmetry, which leads to the simple equation
\citep{sarazin88}:

\begin{equation}
\label{hydro}
M(r) = -4.0\times 10^{13} M_\odot T (keV) \, r (Mpc) \, \Bigg( {{d\log(n_e)}\over{
    d\log r}} + {{d\log(T)}\over{d\log r}}\Bigg) \, .
\end{equation}

Since we are not able to measure the temperature profile, we assume
isothermality inside the extraction radius, while beyond $R_{ext}$ we
adopt a mildly decreasing temperature profile $kT\propto r^{-0.24}$,
as found in local clusters \citep[see][]{leccardi08}.  Therefore the
term in parenthesis in Equation \ref{hydro} is the sum of the slope
of the slowly declining temperature profile and the slope of the
density profile, which is simply $ -3 \, \beta \, x^{2} / (1+x^{2})$,
where $x = r/r_{c}$.

To measure the total mass at a given density contrast, we solve the
equation $M_\Delta(r_\Delta) = 4/3\pi r_\Delta^3 \, \Delta
\rho_c(z_{cl})$.  This relation allows us  to compute the radius where the average density level
with respect to the critical density $\rho_c(z_{cl}) $ is $\Delta$.
Virial mass measurements are typically reported for $\Delta = 2500, 500$,
and $200$.  The 1$\sigma$ confidence intervals on the mass are
computed by including the error on the temperature and on the gas
density profile.  The results are shown in Table \ref{mtab}.  Only the
radius $r_{2500}=240_{-20}^{+30}$ kpc is well within the detection
region.   We obtain $M_{2500} = 1.23_{-0.27}^{+0.46} \times 10^{14}
M_\odot$.  The total mass extrapolated to $r_{500}$, with the temperature profile
gently decreasing beyond $R_{ext}=375$ kpc as $\propto r^{-0.24}$, is $M_{500} =
3.2_{-0.6}^{+0.9} \times 10 ^{14}M_\odot$ for $R_{500} =
562_{-37}^{+50}$ kpc.  We also extrapolate the mass measurement up to
the nominal virial radius, finding $M_{200} = 4.4_{-0.8}^{+1.3} \times
10 ^{14}M_\odot$ for $R_{200} = 845_{-50}^{+80}$ kpc.  The ICM mass
fraction within $R_{ext}$ is  $f_{ICM} = 0.07 \pm 0.02$, and 
$f_{ICM} = 0.08 \pm 0.02$ at $R_{500}$.   These values are
in very good agreement with the empirical relation between
$f_{ICM}$ and $M_{500}$ shown in \citet{2009Vikhlinin}.
In Table \ref{mtab} we also report the mass measured at $R_{ext}=375 $ kpc, 
which is the largest radius where we can obtain a robust measurement of the mass without extrapolations.

\begin{table}
 \centering
\caption{Summary of the mass estimates for XDCP0044 based on the hydrostatic
equilibrium equation \ref{hydro}.  The ``*''
  indicates that the mass values are extrapolated beyond the
  maximum radius $R_{ext}$ where the ICM emission is actually measured.  We assume a constant
  temperature profile within $R_{ext}$ and a slowly decreasing profile as $kT \propto r^{-0.24}$ at larger radii.}
\label{mtab}
\begin{tabular}{@{}cccc@{}}\hline
$\Delta$ & Radius & $M_{tot}$  &     $M_{ICM}$     \\ 
         &  kpc   & $M_\odot$  &     $M_\odot$     \\ 
\hline  \\ 
$2500$  & $R_{2500}=240_{-20}^{+30}$ & $1.23_{-0.27}^{+0.46} \times 10^{14}$ &  $(8.2 \pm 1.1) \times 10^{12} $ \\ 
 \\ 
  -  & $R_{ext}=375$ &  $2.30_{-0.3}^{+0.5}\times 10^{14}$  &    $(1.48\pm 0.2)  \times10^{13} $ \\ 
 \\ 
  $500^*$  & $R_{500} = 562_{-37}^{+50}$ &  $3.2_{-0.6}^{+0.9} \times 10 ^{14}$  &  $(2.5 \pm 0.3) \times 10^{13}$ \\ 
 \\ 
$200^*$  & $R_{200} =  845_{-50}^{+80}$ & $4.4_{-0.8}^{+1.3} \times 10 ^{14}$ &     $(3.8 \pm 0.5) \times 10^{13}$   \\ 
 \\ 
\hline
\end{tabular}
\end{table}

\section{Discussion}

\subsection{Systematics in mass measurements}

The mass measurements at $R_{500}$ and $R_{200}$ have been obtained
under the assumption of hydrostatic equilibrium as an extrapolation of
the observed profile beyond the largest radius where X-ray emission is
detected ($375$ kpc), as well as under some reasonable assumptions on
the slope of the temperature profile declining as $\propto r^{-0.24}$ outside the extraction radius,
as observed in local clusters in the $(0.1-0.6) \times r_{180}$ range
\citep[][]{leccardi08}.  For completeness, we recompute the masses after relaxing the
assumption of isothermality within the extraction radius and adopting
the temperature profile $kT\propto r^{-0.24}$ also for radii $r<R_{ext}$.  In this
case, the only requirement is that the average temperature within $R_{ext}$ 
is $6.7$ keV as observed.  With these assumptions, masses at
$M_{500}$ and $M_{200}$ are $25$\% and $\sim 30$\%, respectively, lower than those obtained with
the simplest choice $T=const$ for $r<R_{ext}$, while $M_{2500}$ is lower
only by 8\%, being much more robust and anchored to the data.  These differences should
be regarded as systematic, since the temperature profile within the
extraction radius may well behave differently from a simple power law,
with significantly different effects on the extrapolated masses.  
In this work, given the uncertain temperature distribution in the ICM of XDCP0044 and the 
spatially resolved spectral analysis presented in \S 2.4, we conclude that the assumption of
constant temperature within $R_{ext}$ and the mass measurements presented in Table \ref{mtab}, 
should be considered as the  most accurate.  Clearly, the uncertainty associated to the temperature profile will be solved only
when a robust, spatially resolved spectral analysis of distant X-ray clusters will be possible.

We also extrapolate the mass according to the Navarro, Frenk \& White profile \citep[NFW]{nfw96}, after requiring a normalization
at 375 kpc to the total mass value actually measured from the data.  Still, the extrapolation depends on the unknown concentration
parameter.  However, such dependence is not strong and can be accounted for once we restrain the concentration in the
plausible range $c_{NFW} = 4.0 \pm 0.5$ \citep{gao08} or $c_{NFW} = 3.5 \pm 0.5$ \citep{2008Duffy}, as found in simulations for
the wide mass range $5\times 10^{13} M_\odot< M < 5\times 10^{14}M_\odot$ at high $z$.  We find $M_{500} = (3.7_{-0.9}^{+1.5})
\times 10^{14}M_\odot$ and $M_{200} = (5.6_{-1.7}^{+1.9}) \times 10^{14}M_\odot$.  Here, the much larger error bars include both the
propagation of the statistical error on the measured mass within $R_{ext}$, which is used to normalize the NFW profile, and the uncertainty on the concentration parameter.
 
\subsection{Comparison with empirical calibrations}

\begin{table}
 \centering
\caption{Total mass estimates of XDCP0044 at $R_{500}$ and $R_{200}$ obtained through empirical calibrations.  (1) \citet{2009Vikhlinin}; (2) \citet{2011Reichert}; (3)
  \citet{2009Vikhlinin}; (4) \citet{fabjan11}.}
\label{mtab_extrap}
\begin{tabular}{@{}cccc@{}}\hline
Method & $R_{500}$    & $M_{500}$   & $M_{200}$   \\ 
             & kpc             & $M_\odot$   & $M_\odot$   \\ 
\hline  \\ 
M-T relation (1) & - & $ 2.8^{+0.8}_{-0.6} \times 10^{14} $ & -  \\ 
 \\ 
M-T relation (2) & - & - & $ 4.0^{+1.7}_{-1.4} \times 10^{14} $  \\ 
 \\ 
$Y_X-M$ relation  (3) & $515 \pm 30$  & $2.2_{-0.4}^{+0.5} \times 10^{14} M_\odot$  &  - \\
 \\ 
$Y_X-M$ relation  (4) &  $580_{-35}^{+45} $ &  $ 3.3_{-0.6}^{+0.4} \times 10^{14} M_\odot$ & - \\
 \\ 
\hline
\end{tabular}
\end{table}

In order to evaluate the uncertainties on the estimate of  $M_{500}$ and $M_{200}$ based on empirical relations, we present the mass estimates adopting four different
calibrations. Clearly, here we are using the high-z end of these calibrations, which have been obtained mostly on the basis of low and medium-z clusters.  
The first  estimate is based on the redshift-dependent scaling relations calibrated on local clusters and presented in
\citet{2009Vikhlinin}.  From the empirical relation described in their Table 3:

\begin{equation}
M_{500} =M_0 \times ( kT/ 5 \, {\tt keV})^\alpha E(z)^{-1} \, ,
\end{equation}

\noindent
where $M_0 = (2.95 \pm 0.10) \times 10^{14} h^{-1} M_\odot$ and $\alpha = 1.5$.
We find $M_{500} = 2.8^{+0.8}_{-0.6} \times 10^{14} M_\odot$, consistent well within $1\sigma$ with our
measurement based on the hydrostatic equilibrium equation.   Another mass-temperature calibration, based on distant
cluster measurements \citep{2011Reichert}, provides an estimate for the virial mass $M_{200} = 4.0_{-1.4}^{+1.7} \times 10^{14} \, M_\odot$.

As a further method, we use the integrated pseudo-pressure parameter $Y_X \equiv T_X \times M_{ICM}$, which is considered a robust mass proxy within
$R_{500}$, as shown by numerical simulations \citep[e.g.,][]{kravtsov06}.  The observed value for XDCP004  is $Y_X = (1.60 \pm 0.33)\times 10^{14}$ keV $M_\odot$.  
We use the $Y_X$-$M$ empirical relation taken from Table 3 of \citet{2009Vikhlinin}:

\begin{equation}
M_{500} = M_0 \times (Y_X/3\times 10^{14} M_\odot \, {\tt keV})^{\alpha} E(z)^{-2/5}  \, ,
\end{equation} 

\noindent
where $M_0 = (2.95 \pm 0.30) \times  10^{14} M_\odot h^{1/2}$ and $\alpha = 0.6$. 
We  find $M_{500} = (2.2_{-0.4}^{+0.5}) \times 10^{14} M_\odot$ and $R_{500} = (515 \pm 30)$ kpc.
Using the calibration based on numerical simulations obtained in \citet{fabjan11}, we find $M_{500} = 3.3_{-0.6}^{+0.4} \times 10^{14}
M_\odot$, for $R_{500} = 580_{-35}^{+45} $ kpc.  Both values are obtained iteratively in order to compute consistently all the quantities at $r_{500}$.  
Note, however, that $Y_X$ should be measured between $0.15 \times R_{500}$ and $R_{500}$ and $M_{gas}$ 
within $R_{500}$, while for XDCP0044 we measure the global temperature within $R_{ext}<R_{500}$ and $M_{ICM}$ is computed
at $R_{500}$ by extrapolating the best-fit $\beta$ model.   Therefore, our measured value of $Y_X$ may differ 
from the true pseudo-pressure parameter.  We also comment that the use of $Y_X$ as a mass proxy at high redshift may
require higher S/N.

The uncertainty among different calibrations can be appreciated in Table \ref{mtab_extrap}.  We find that
the uncertainty on $M_{500}$ can amount to 20\%, comparable to the $1 \sigma$ confidence level associated to the statistical
uncertainty.  We remark that mass estimates based on a $\beta$-profile fitting and the assumption of the hydrostatic
equilibrium may give values $\sim 20$\% lower around $r_{500}$ (and even more discrepant for larger radii) due to the violation of hydrostatic
equilibrim and the presence of significant bulk motions in the ICM. This has been shown in numerical simulations
\citep[see][]{bartelman96,rasia04,borgani04}.  Observational calibrations also show that hydrostatic masses underestimate weak
lensing masses by 10\% on average at $r_{500}$ \citep[see][]{hoekstra07,2013Mahdavi}.  This is also found in numerical simulations \citep[see][]{2011Becker,2012Rasia}.  
However, most recent works \citep{2014vonderLinden,2014Israel} find no hints of bias, at least in massive clusters, by comparing X-ray and weak lensing masses.  

Our conclusion is that the use of empirical calibrations to estimate $M_{500}$ may suffer systematics  of the order of 20\%, comparable, in this case, to the statistical errors associated to the 
X-ray mass measurements.  Estimates of $M_{200}$ are more uncertain (30\%-40\%).  A weak lensing study will be crucial to assess
the robustness of the X-ray mass estimate at $R_{200}$ for this cluster.
We expect this to be feasible with moderately deep HST-ACS imaging, based on \citet{jee11}.

\subsection{Cosmological implications}

Although several clusters have been confirmed at $z\gtrsim 1.5$, currently XDCP0044 is the only cluster  at these redshifts
whose diffuse X-ray emission provides constraints on the ICM temperature
and on the mass.  Because the expected number of hot clusters at such high redshift is extremely low, 
the mere presence of a few massive, high-z galaxy clusters could
create some tension with the standard $\Lambda$CDM and quintessence
models \citep[e.g.,][]{jimenez09,baldi11,chong12,2011Mortonson,2012Harrison,2012Waizmann}.
So far, no evidence for such a discrepancy has been found, except when
considering the combined probability of the most massive, high-z clusters \citep{jee11}.
Therefore, it is important to investigate whether the presence of XDCP0044 is creating tension with the
current $\Lambda$CDM paradigm of the large-scale structure formation.

It is straightforward to predict the abundance of massive clusters at $z>z_{min}$ whose mass exceeds 
$M>M_{min}$ by evaluating the following integral:
\begin{equation}
N(M,z) = f_{sky} \int_{z_{min}}^{z_{max}}  \frac{dV(z)}{dz} dz \int_{M_{min}}^{M_{max}}  \frac{dn}{dM} dM \, ,
\label{eqn_abundance}
\end{equation}
\noindent
where $dV/dz$ is the volume element per redshift interval, $dn/dM$ is the mass function, and $f_{sky}$ 
is the survey area normalized to give $f_{sky}=1$ for a full sky.  In general, $f_{sky}$ depends on the flux
and therefore on the cluster mass and redshift, so that $N(M,z)$ would depend on the convolution of $f_{sky}$ with the 
mass function and the volume element.  However, at present, we  cannot  model $f_{sky}$ for the XDCP survey
and we choose a constant, conservative value as explained below.

In this paper, we use the \citet{2008Tinker} mass function, which has been calibrated based on numerical simulations.  We do not attempt to include the effects of baryon 
physics, which have been shown to be relevant also at cluster scales.  In particular, it has been found that AGN feedback produces a lower mass function with 
respect to the DM-only case \citep{2014Velliscig,2014Cui}.  Clearly, this effect would reduce the probability of finding massive clusters for a given set of cosmological 
parameters.  Therefore, our choice of adopting the \citet{2008Tinker} mass function must be considered as conservative.  Finally, because the cluster X-ray emission is 
mostly confined within a few hundred kpc from the center, we evaluate the mass function using a contrast of $\Delta=600$ with respect to 
the critical density, which gives the cluster mass $M_{600} \simeq 3.1\times10^{14} M_{\Sun}$ at $r_{600} = 530$ kpc.

The median flux limit of the XDCP survey is $\mytilde 10^{-14}$ $\mbox{erg}~\mbox{s}^{-1}~\mbox{cm}^{-2}$
in the 0.5-2 keV band, which we translate into a maximum detection redshift $z_{max}\sim 2.2$. We remind 
readers that the rarity of massive clusters at high redshift makes the evaluation of the integral not sensitive to the 
exact value of the redshift upper limit.  For this flux limit, the best estimate of the effective survey area is $\mytilde 70$ sq. deg ($f_{sky} \sim 1.7 \times 10^{-3}$).
We stress that this choice corresponds to the assumption that the XDCP reaches a limiting flux of $10^{-14}$ $\mbox{erg}~\mbox{s}^{-1}~\mbox{cm}^{-2}$
over the entire solid angle.  A more realistic treatment would give a lower $f_{sky}$, and correspondingly lower probabilities of finding such a massive cluster in the XDCP.  
Finally, the probability distribution is computed by setting our cluster's redshift as the minimum value, therefore  we are not considering the
degeneracy between the $M$ and $z$ values for a given probability \citep[see][]{2011Hotchkiss}.

Adopting the above $M_{600}$ as our threshold mass $M_{min}$, we find that the probability of 
discovering at least one cluster with mass larger than $M_{min}$ and redshift higher than $z_{min}$ within the XDCP survey volume  is $\mytilde 3$\% and $\mytilde 6$\% using the central 
values of the WMAP and PLANCK cosmological parameters, respectively. An interesting question is 
whether or not we should interpret this probability as indicating any tension with our $\Lambda$CDM paradigm. 
Obviously, in order to adequately propagate the discovery probability into cosmological tension, 
the following additional factors should be considered. 

First, we need to take into account the statistical uncertainties of current cosmological parameter measurements. 
For example, if we increase the $\sigma_8$ value by 2$\sigma$ ($\sigma_8=0.882)$,  the probability becomes as high as $\mytilde16$\%.
Second, an Eddington bias may be a non-negligible factor for objects with a steep mass function. It is possible that our 
central value of $kT=6.7 $ keV may be obtained by an up-scatter while the ``true'' temperature of the cluster is lower or 
that our result is given by a down-scatter from a higher temperature. Because of the steep mass function, the up-scatter is more likely than the down-scatter. 
\citet{2011Mortonson} suggested that one should use a lower (corrected) mass to properly compensate for the Eddington bias. 
Substituting the central value of our mass estimate into Equation 16 of  \citet{2011Mortonson}  reduces the above 
threshold mass by $\mytilde 5$\%. Therefore, in the current case the application of  
this correction does not critically affect our cluster abundance estimation.
Third, one can question the validity of adopting the cluster redshift $z=1.58$ as $z_{min}$. Using the cluster redshift is 
justified when we merely estimate the expected abundance of a similarly massive cluster beyond that redshift. 
However, for examining cosmological implications, one cannot exclude the volume at lower redshift, where more massive 
clusters with similar rarity can exist. Estimating a fair value for $z_{min}$ is challenging in general 
(e.g., priority in follow-up observations) and particularly so in the XDCP survey where observers 
preferentially look for faint, distant cluster emission in the archival data. 
Fourth, we should remember that current theoretical mass functions are obtained in 
$N$-body simulations, where extreme clusters are also rare.  Consequently, the mass function is the result of extrapolation.
Fifth, the current cluster mass is derived under the hydrostatic equilibrium assumption. 
Although we have witnessed a case where the assumption may still hold for massive clusters such as 
XMMUJ2235 at $z=1.4$ \citep{2009Rosati,2009Jee},  the redshift $z=1.58$ is a regime where we expect a rapid evolution of ICM properties. 

Considering the above factors, we conclude that it is premature to translate the rarity of XDCP0044 by itself into any tension with 
the current $\Lambda$CDM paradigm. However, it is still interesting to note that the relatively small XDCP survey 
have already discovered two extreme clusters: XDCP0044 and XMMUJ2235 at z=1.58 and 1.4, respectively. 
The  smaller ($\mytilde8$ sq. deg) IDCS survey found the cluster IDCS J1426.5+3508 at $z=1.75$, whose discovery 
probability within that survey is $\mytilde 1$\% \citep{2012Brodwin}. The much larger (1000 sq. deg) 
ACT SZ survey discovered the ``El Gordo'' cluster $z=0.87$ \citep{2012Menanteau} whose X-ray and weak-lensing masses 
indicate that the probability of discovering such a massive cluster is low ($\mytilde 1$\%) even in the full sky 
\citep{2013bJee}. To summarize, there are hints that the detection of sparse, massive
galaxy clusters at high redshift points toward inconsistencies of the standard $\Lambda$CDM  model or a significantly higher normalization ($\sigma_8$)
parameter.  However, more robust studies based on deep and
complete samples of galaxy clusters are necessary to quantify the actual tension with the $\Lambda$CDM.  In this respect, 
SZ surveys, coupled to a robust mass calibration, may play a dominant role.
Unfortunately, it is very hard to build deep and complete samples with current X-ray facilities, and the planned 
X-ray surveys in the near future will not be able to explore efficiently the high-z range (see also \S 4.5).

\subsection{Substructure and dynamical status}

A visual inspection of the residual ICM emission shows a clear elongation along the North-South axis.  We already investigated
in \S 2.4 and \S 2.5 the possible existence of two clumps with different temperatures. 
We showed that both the spectral analysis and the azimuthally averaged surface brightness distribution 
suggest the presence of two distinct clumps constituting the bulk of the ICM.  However, deeper data
would be needed before reaching firm conclusions.  In fact, it is impossible to decide whether
the elongation and the two-region structure  are due to an ongoing merger between comparable 
mass halos or simply to a young dynamical status.   If the typical formation epoch of massive
clusters is $z\leq 2$, the age of XDCP0044 is less than 0.8 Gyr, 
a time scale which is shorter than its  dynamical time.  In this scenario, we expect
to observe significant differences in virialized, massive clusters in the $1.5 <z<2.0$ range, since the time 
elapsed from the first virialization process is smaller than their dynamical time scales.

\begin{figure}
\includegraphics[width=7cm]{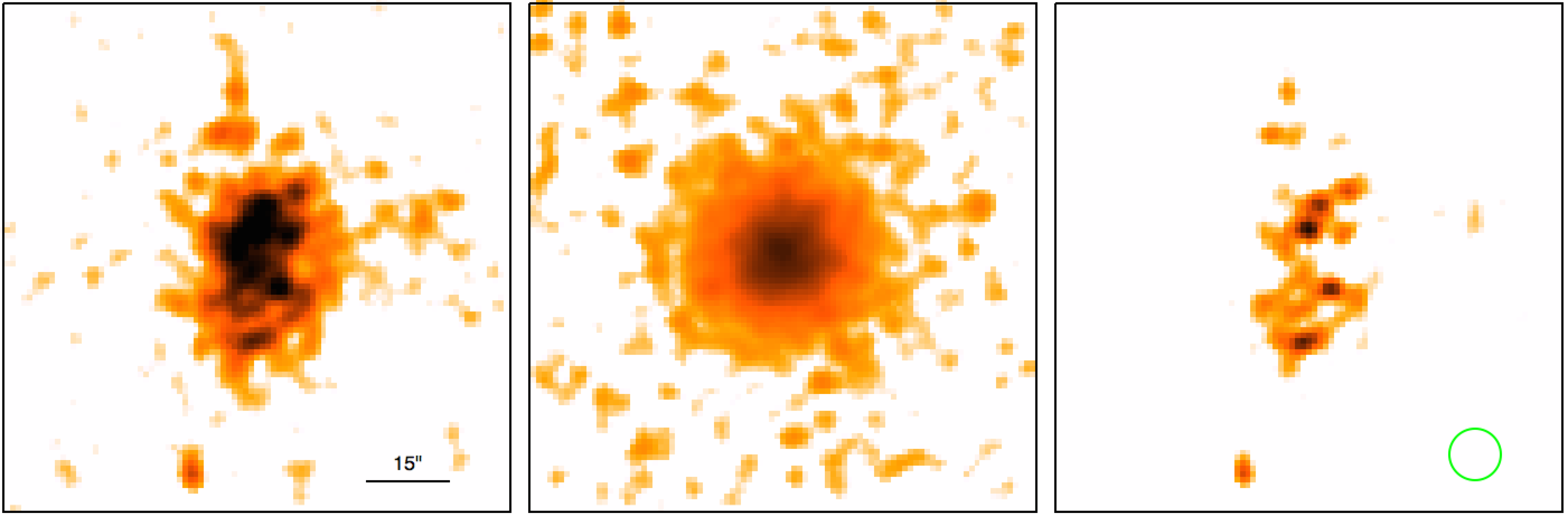}
\caption{\label{residual}Left panel: smoothed X-ray image of XDCP0044 in the soft (0.5- 2 keV) band after removing the identified unresolved sources.  Central panel: 
Simulated soft-band image from the best-fit $\beta$ model.  Right panel: map of the X-ray residual in the soft  band obtained subtracting the simulated
$\beta$-model image from the X-ray image.  All images are $90\arcsec$ across.  The circle in the lower right of the third panel corresponds to a radius of
$\sim 40$ kpc, which is the scale where we found  evidence of substructures in the ICM.}
\end{figure}

To further investigate the presence of substructures in the ICM distribution, we remove the best fit $\beta$-model from the X-ray image to obtain the residual image.  In Figure
\ref{residual} we show the smoothed X-ray image of XDCP0044 in the soft band, a simulated image of the best fit model and the residual image, obtained subtracting an average of $10^4$ simulated 
images from the real data.  The field of view of the three images is a square with a side of $\sim 90"$, corresponding to 180 pixels (1 pixel corresponds to $0.492"$).  
In the smoothed image, the surface brightness distribution appears to be  clumpy along the direction of the elongation, and this appears more clearly in the residual image.  
However, the statistical  significance of the clumps must be carefully evaluated before drawing any conclusion.

\begin{figure}
\begin{center}
\includegraphics[width=15cm]{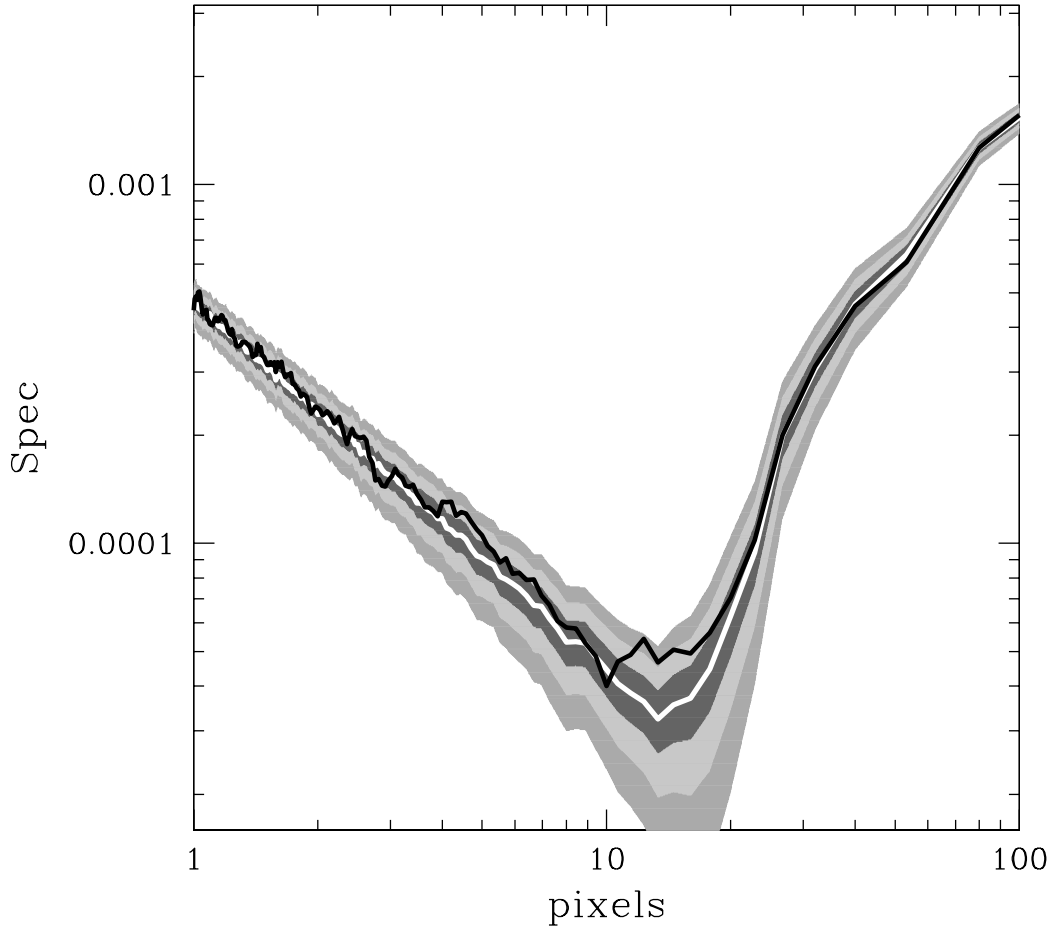}
\caption{\label{spectra_XDCP0044}The radially averaged power spectrum of the soft-band image of XDCP0044 (black solid line) is compared to the expected power spectrum of the 
image of the best-fit $\beta$-model, obtained as an average over $10^4$ simulated images.  The power spectrum is shown as a function of the physical scale (in pixels).  The size of the each 
simulated image is $320 \times 320$ pixels, corresponding to $157.4"$.  The shaded areas corresponds to 1, 2 and 3 $\sigma$ confidence (from dark to light grey).  
We notice a $\sim 2 \sigma$ excess in the data with respect to the model for scales between 10 and 20 pixels.  } 
\end{center}
 \end{figure}

To assess the significance of the fluctuations in the X-ray surface brightness of XDCP0044, we proceed as follows.  First we create $\sim\! 10000$ X-ray images
obtained directly as random realizations from the best fit $\beta$-model.  We use the same background and the same number of total counts from the real soft band
image, in order to reproduce the same level of noise in the data.  We also convolve the simulated images with the normalized, soft-band exposure map, in order to take into account any possible feature
due to variations of the effective area.  This set of images can be compared with the XDCP0044 soft-band image, where all the unresolved sources previously identified
have been  removed  and the corresponding extraction regions are filled with photons, consistently with the surrounding surface brightness level.  We verify {\sl a-posteriori} that this procedure 
does not introduce artificial features in the XDCP0044 image.  As a simple but effective estimate of the amount of substructures in the simulated images, 
we compute the power spectrum as a function of $| K |$, the magnitude of the wavenumber vector. This is accomplished by computing the Fast Fourier Transform of each image 
and averaging power over circles of diameter $| K |$.  We choose an image size of $320 \times 320$ pixels, corresponding to $157.4"$.  The expected spectrum 
from the best-fit $\beta$-model is computed averaging over the $10^4$ simulated images.  The crucial information is not only the average power spectrum, but also its variance 
due to the actual S/N of the images.   In Figure \ref{spectra_XDCP0044} we show the average power spectrum of the simulated 
images and its variance.  We also compute the 2D average power spectrum of the real image of XDCP0044, which is also shown in Figure \ref{spectra_XDCP0044} as a black continuous line.
We notice a $\sim 2 \sigma$ excess in the data with respect to the model corresponding to scales between  10 and 20 pixels.  

\begin{figure}
\begin{center}
\includegraphics[width=15cm]{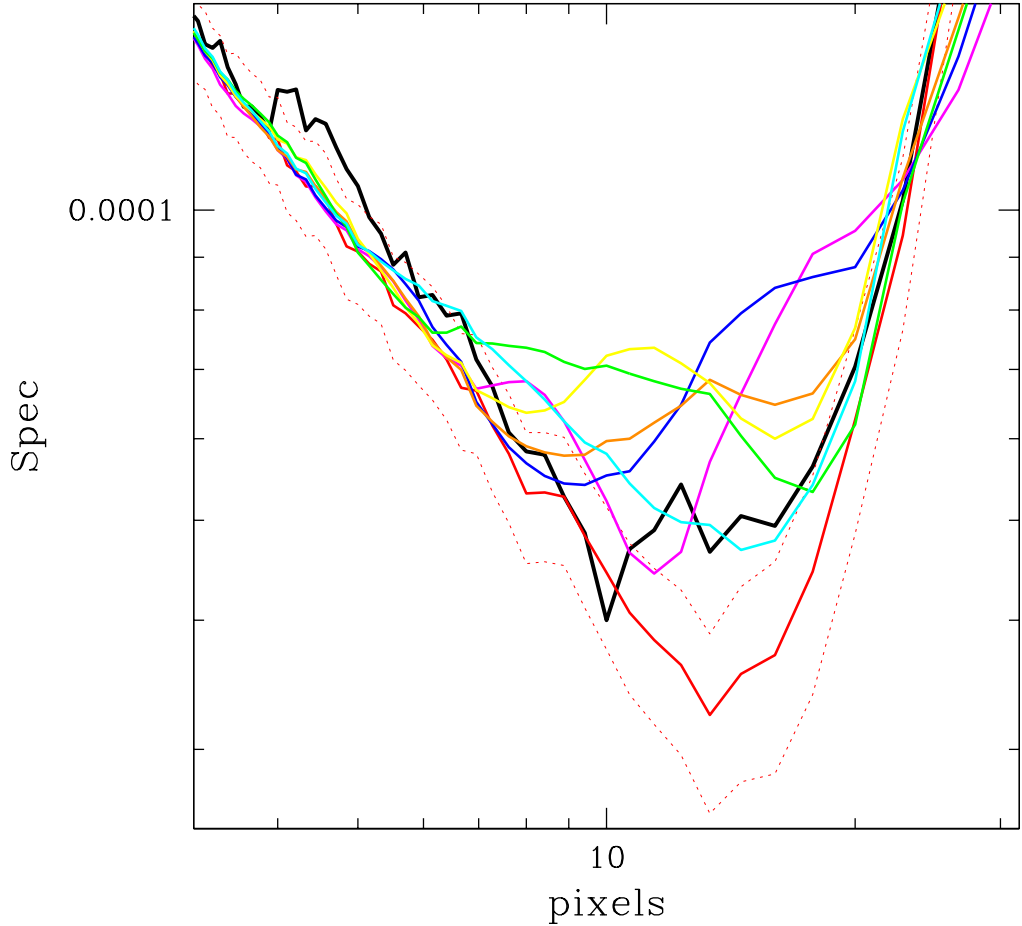}
\caption{\label{spectra_check}The radially averaged power spectrum of the soft-band image of XDCP0044 (black solid line) is compared to the expected power spectrum of the 
image of the best-fit $\beta$-model with top-hat perturbations with a radius of 5, 7, 8, 9 10 and 15 pixels, shown as  cyan, yellow, green, orange, blue and  magenta lines, respectively.
The average power spectrum of the best-fit images and its 1 $\sigma$ uncertainty are shown with red solid and dashed lines, respectively.} 
\end{center}
 \end{figure}

In order to better understand this result, we use another set of simulations, obtained by adding holes and bumps to the best-fit $\beta$-model. The holes and bumps are modeled as circular, randomly placed 
top-hat cavities or enhancements in the surface brightness distribution.  These substructures are concentrated towards the cluster center, and only a limited number of configurations are simulated, 
which limit our investigation of possible substructures in the surface brightness of the ICM.  However, our procedure is adequate for a first characterization of the typical scale of the substructures.  
The results are shown in Figure \ref{spectra_check}, where we show the color-coded average power spectra of the simulated images with top-hat perturbations with a radius of 5, 7, 8, 9, 10 and 15 pixels.  
For clarity, the amplitude of the artificial perturbations  in the simulations is larger than that expected in real data. 
As already mentioned, the size, position and number of top-hat perturbations give rise to different spectral shape.  Therefore, it is not straightforward to identify the typical 
perturbation scale directly from a visual inspection of the 2D averaged power spectrum.  However, a clear trend is observed from the smallest to the largest scales.   
A comparison with the spectrum of the real data, suggests that the $\sim 2 \sigma$ excess present in the power spectrum of the XDCP0044 image can be ascribed to roughly circular
fluctuations with radius of $\sim 9$ pixels, corresponding to a physical scale of about 40 kpc.  In the third panel of Figure \ref{residual} we show a circle with a 40 kpc radius for a direct comparison
with the residual image.

In summary, we find $2 \sigma$ evidence of substructures in the surface brightness distribution of XDCP0044, compatible with circular, top-hat perturbations with a typical radius of 40 kpc.  
This can be interpreted with the presence of ICM clumps associated to relatively recent merging, but also with the presence of cavities.
We stress that, despite the relatively low S/N of high-z clusters’ observations, a statistical analysis of their surface brightness
fluctuations is still possible with {\it Chandra}.  The extension of this analysis to the overall high-z cluster population will be presented elsewhere.

\subsection{Future surveys and high-z clusters}

Clusters of galaxies at high redshift ($z>1$) are vitally important to understand 
the evolution of the large scale structure of the Universe, the processes 
shaping galaxy populations and the cycle of the cosmic baryons, and to constrain cosmological parameters.  
At present, we know that at $z>1$ many massive clusters are fully virialized and
their ICM is already enriched with metals.  The present study extends this picture to $z\sim 1.6$.
In the near future, it is possible that new studies will reveal virialized clusters at larger redshift and more massive than XDCP0044 \citep[for example IDCS1426 at $z=1.75$, which has been awarded a
deep Chandra exposure in GO 14,][]{2012Stanford}

Clearly, the small number statistics prevents us to draw quantitative conclusions on the
evolutionary behaviour in the range $1<z<1.6$, namely the evolution of the iron abundance and the evolution of the cool core phenomenon.  
The assembly of a large and well characterized sample of high-z X-ray clusters is a major goal
for the future.  For a basic  characterization of an X-ray cluster, we consider the collection of about 1500 net photons 
in the 0.5-7 keV band with high angular resolution (point spread function HEW $\sim 1-2"$) images to remove the
effect of contaminating AGN emission and identify central cool cores.  These requirements can provide a robust 
measurement of $M_{2500}$ and a reliable estimate of $M_{500}$, necessary to perform cosmological tests.  

The present study provides the first characterization of a massive cluster at a redshift as high as $1.6$, but, at the same time, 
clearly shows that the realm of high-z clusters ($z\sim 1.5$ and higher,
when the lookback time is larger than 9 Gyr) requires very time-expensive observations even with 
a major X-ray facility like {\sl Chandra}.  At present only {\sl Chandra} has the angular resolution 
needed to achieve an in-depth analysis of the ICM properties of distant clusters.  In the near future it is likely that the number of detections of high-z X-ray clusters will increase
to several tens on the basis of the still growing {\sl Chandra} and XMM-Newton archives.
However, the number of high-z X-ray clusters with a robust physical characterization and measured hydrostatic masses will be much lower given its very high cost in terms of observing time.

The perspective for distant cluster studies is not expected to improve much on the basis of currently planned missions.
Looking at the near future, the upcoming mission eROSITA \citep{2010Predehl,2012Merloni}  will finally provide an X-ray all-sky
coverage 20 years after the ROSAT All Sky Survey \citep{1999Voges},  down to limiting fluxes
more than one order of magnitude lower than ROSAT for extended sources, therefore considerably increasing the number of 
low and moderate redshift clusters.   However, the limiting flux after
four years of operation is predicted to be $3.4 \times 10^{-14}$ erg s$^{-1}$ cm$^{-2}$, well above the $10^{-14}$ erg s$^{-1}$ cm$^{-2}$
flux level below which  the majority of the distant cluster population lie.  A sensitivity level of $10^{-14}$ erg s$^{-1}$ cm$^{-2}$ will be reached only in the pole regions on $\sim 140$
deg$^2$, a solid angle which is already covered by the extragalactic {\sl Chandra} archive at better fluxes and much better angular
resolution.  Therefore, in the near future the number of new high-redshift clusters will be mainly provided by SZ observations and
other near-IR large-area surveys, some of which are already delivering an increasing number of clusters at $z > 1$
\citep[see][]{2014Rettura}.  However, X-ray observations will still be required for a physical characterization of these systems, mass
calibration and a wide range of astrophysical and cosmological applications.

Only a wide-field, high angular resolution X-ray mission with a large
collecting area and good spectral resolution up to 7 keV seems to be
able to match such requirements \citep[see the WFXT,][]{2010Murray}.
The combination of good angular resolution and a constant image
quality across a 1 deg$^2$ FOV, coupled with a large effective area,
and a survey-oriented strategy, can provide a direct measurement of
temperatures, density profiles and redshifts for a least 1000 if not
several thousands X-ray clusters at $z>1$ , depending on the survey
strategy.  We stress that the capability to perform detailed studies
for $z>1$ clusters critically depends on the angular resolution, to
avoid the confusion limit down to fluxes much lower than $10^{-14}$
erg s$^{-1}$ cm$^{-2}$.  In addition, the capability of measuring the X-ray redshift directly
from the X-ray analysis of the ICM emission would avoid a
time-prohibitive optical spectroscopic follow-up program.  This kind
of mission would improve by almost two orders of magnitude any
well-characterized cluster sample that we can possibly assemble using
the entire wealth of data from present and planned X-ray facilities.

In the distant future (15-20 years) major X-ray missions like Athena \citep{2012Barcons,2013Nandra} and SMART-X can provide surveys with the required
depth and quality, by devoting a significant amount of their lifetime
to such a program.  In particular, if used in survey-mode, the current design of SMART-X
with a large FOV CMOS detector would be comparable with WFXT.  Clearly, the 
characteristics of SMART-X are best suited for a follow-up campaign of, let's say, SZ-selected clusters, 
rather than for an X-ray survey.  In general, given their optics and their instrument
setup, it is not very efficient to use these two major X-ray
facilities in survey mode.  Also, Athena or SMART-X surveys will be
available much later than the many wide surveys that will dominate
Galactic and extragalactic astronomy in the next decade (Pan-STARRS,
LSST, Euclid, JVLA, SKA).  To summarize, despite eROSITA will provide a crucial
and vital all-sky survey in the soft X-ray band, all the scientific
cases concerning objects in the faint flux regime (about $10^{-14}$
erg cm$^{-2}$ s$^{-1}$ and below), or requiring substantial
information above 2 keV, has to rely on archival data of {\sl Chandra} and XMM-Newton.

\section{Conclusions}

We presented a deep X-ray observation of the most massive distant
X-ray cluster of galaxies presently known.  XDCP0044 was discovered in
the XDCP survey and imaged in a 380 ks {\sl Chandra} ACIS-S exposure
aimed at measuring its mass and studying its ICM properties. We are
able to obtain a robust measure of the ICM temperature and total mass
within 375 kpc and a reliable extrapolation up to $R_{500}$.  We are
also able to investigate the presence of substructures in the ICM
distribution.  Here we summarize our main results:

\begin{itemize}

\item XDCP0044 is detected with a $S/N\sim 20$ within a circle with a radius of $44"$, corresponding to $375$ kpc at $z=1.58$.  
The azimuthally averaged surface brightness distribution can be well
described by a $\beta$ model with $\beta = 0.75$ and $r_c = 99$ kpc; the soft-band flux within
  $r=44\arcsec$ is equal to $S_{0.5-2.0 keV}=(1.66 \pm 0.01) \times 10^{-14}$ erg s$^{-1}$ cm$^{-2}$;

\item the spectral fit with a {\tt mekal} model gives a global temperature
  $kT=6.7^{+1.3}_{-0.9}$ keV and a global iron abundance $Z_{Fe} =
  0.41_{-0.26}^{+0.29}Z_\odot$;

\item despite the  iron emission line is measured with a confidence level lower than 2$\sigma$,
the upper limits found for the iron abundance are in agreement with a mild, negative evolution of about a factor 1.5-2 between $z=0$ and $z\sim 1$ as found in previous Chandra and XMM-Newton studies
\citep{2007Balestra,2008Maughan, 2009Anderson,2012Baldi}, and, 
at the same time, with a non-negligible chemical enrichment at a level of $\sim
0.2-0.4 Z_{Fe_\odot}$, comparable to local clusters;  

\item the total mass measured at $R_{2500}=(240_{-20}^{+30})$ kpc is $M_{2500} =
  1.23_{-0.27}^{+0.46} \times 10^{14} M_\odot$.  The total mass is
 reliably  measured out to $375$ kpc, where it amounts to $M(r<375 \, \, {\tt kpc}) =
  2.30_{-0.30}^{+0.50} \times 10 ^{14}M_\odot$, while the ICM mass is
  $M_{ICM} (r<375 {\tt kpc}) =(1.48 \pm 0.2 ) \times 10^{13}M_\odot$,
  resulting in a ICM mass fraction $f_{ICM} = 0.07\pm 0.02$;

\item  assuming a mildly decreasing temperature profile at $r>R_{ext}$ as
  observed in local clusters, we are able extrapolate the mass
  measurement up to $R_{500} = 562_{-37}^{+50}$ kpc, finding $M_{500} =
  3.2_{-0.6}^{+0.9} \times 10 ^{14}M_\odot$; this value is consistent
  with those obtained using empirical relations between X-ray
  observables and total mass;

\item a detailed analysis of the surface brightness distribution reveals the presence of
two clumps, with marginally different temperatures, which may be due to a recent merger or
to a young dynamical status; we do not find evidence for the presence of a cool core; 
the average values of the entropy $\sim 94^{+19}_{-14}$ keV cm$^{-2}$ and of the cooling time 
$4.0^{+0.9}_{-0.6}$ Gyr within 100 kpc confirm the lack of a cool core;

\item we also find some evidence ($2 \sigma$ c.l.) for clumping in the surface brightness distribution on scales $\sim 40$ kpc, 
which may be interpreted as the signature of a not fully relaxed dynamical status, possibly due to 
the young age of the cluster;

\item when compared with the expectations for a $\Lambda$CDM universe
  based on the mass function of \citet{2008Tinker}, XDCP0044 appears to
  be a typical cluster at $z\sim 1.6$ for a WMAP cosmology
  \citep{2011Komatsu}; however, the redshift and the total mass of
  XDCP0044 make it the most massive galaxy cluster identified at
  $z>1.5$.

\end{itemize}

The comparison of the ICM properties and of the galaxy population in this cluster, which will be presented in a series of
future papers, may cast new light on the formation epoch of massive clusters, when processes like chemical enrichment, feedback from AGN, 
induced starbursts in the member galaxies, and merger events combine together in a short but hectic epoch.  A systematic study of the distant cluster population, which is 
currently beyond the capability of the present-day X-ray facilities without a prohibitives investment of observing time, will reveal a huge wealth of complex processes whose comprehension is 
mandatory for a complete picture of cosmic structure formation.

\acknowledgements

We acknowledge financial contribution from contract PRIN INAF
2012 ({\sl "A unique dataset to address the most compelling open questions about X-ray galaxy clusters"}), 
and PRIN MIUR 2009 ({\sl “Tracing the growth of structures in the Universe”}). 
H.B. acknowledges support from the DFG Transregio program 'Dark Universe' and the Munich Excellence Cluster
'Structure and Evolution of the Universe'.  This work was carried out with \textit{Chandra} Program 14800360 obtained in AO14.  
JSS acknowledges funding from the European Union Seventh Framework  Programme (FP7/2007-2013) under grant
agreement number 267251 Astronomy Fellowships in Italy (AstroFIt).
 We acknowledge the hospitality of the Villa Il Gioiello (Arcetri, Florence), which is part of the agreement 
regarding the {\sl Colle di Galileo}, where some of the results presented here were discussed.
Finally, we thank the referee, Pasquale Mazzotta, for his careful and critical reading of this manuscript.

\bibliography{references_XMMUJ0044}

\end{document}